\documentclass[10pt]{amsart}
\usepackage{amsmath}
\usepackage{latexsym}
\usepackage[dvips]{graphicx}

\newtheorem{thm}{Theorem}[section]
\newtheorem{lem}[thm]{Lemma}
\newtheorem{prop}[thm]{Proposition}
\newtheorem{cor}[thm]{Corollary}

\theoremstyle{definition}

\newtheorem{axiom}{Axiom}[section]

\def\id{\operatorname{id}}

\def\veck{{\text{\boldmath$k$}}}

\def\vecm{{\text{\boldmath$m$}}}
\def\vecn{{\text{\boldmath$n$}}}

\def\vecpsi{{\text{\boldmath$\psi$}}}

\def\vecxi{{\text{\boldmath$\xi$}}}

\newcommand{\lcal}{\mathcal{L}}
\newcommand{\mcal}{\mathcal{M}}

\def\vecA{{\text{\boldmath$A$}}}
\def\vecB{{\text{\boldmath$B$}}}
\def\vecH{{\text{\boldmath$H$}}}
\def\vecI{{\text{\boldmath$I$}}}
\def\vecJ{{\text{\boldmath$J$}}}

\def\vecU{{\text{\boldmath$U$}}}

\def\scrB{{\mathcal B}}

\def\scrD{{\mathcal D}}
\def\scrF{{\mathcal F}}

\def\scrI{{\mathcal I}}
\def\scrK{{\mathcal K}}

\def\scrM{{\mathcal M}}

\def\scrS{{\mathcal S}}

\def\scrU{{\mathcal U}}
\def\scrV{{\mathcal V}}

\def\CC{{\mathbb C}}

\def\NN{{\mathbb N}}
\def\QQ{{\mathbb Q}}
\def\RR{{\mathbb R}}
\def\SS{{\mathbb S}}
\def\TT{{\mathbb T}}
\def\ZZ{{\mathbb Z}}

\def\vecnull{{\text{\boldmath$0$}}}

\def\fraM{{\mathfrak M}}

\def\e{\mathrm{e}}
\def\i{\mathrm{i}}

\def\const{\operatorname{const}}

\def\id{\operatorname{id}}

\def\C{\operatorname{C{}}}

\def\L{\operatorname{L{}}}
\def\M{\operatorname{M{}}}
\def\R{\operatorname{R{}}}

\def\Op{\operatorname{Op}}
\def\vecOp{{\text{\boldmath$\operatorname{Op}$}}}

\def\sym{\operatorname{sym}}

\def\supp{\operatorname{supp}}

\def\Tr{\operatorname{Tr}}

\def\sign{\operatorname{sign}}

\def\schi{\widetilde{\chi}}

\numberwithin{equation}{section}

\begin{document}

\title[Weyl's law and quantum ergodicity]
{Weyl's law and quantum ergodicity for maps with divided phase space}
\author[Jens Marklof and Stephen O'Keefe]{By Jens Marklof and Stephen O'Keefe}
\address{Jens Marklof and Stephen O'Keefe,
School of Mathematics, University of Bristol,
Bristol BS8 1TW, United Kingdom}
\address{Steve Zelditch, Department of Mathematics,  Johns Hopkins University,
Baltimore, MD 21218,
U.S.A.}
\address{{\tt j.marklof@bristol.ac.uk}}
\address{{\tt stephen.okeefe@bristol.ac.uk}}
\address{{\tt szelditch@jhu.edu}}

\thanks{Research of J.M. supported by an EPSRC Advanced Research Fellowship,
EPSRC Research Grant GR/R67279/01,
Royal Society Research Grant 22355, and the
EC Research Training Network (Mathematical Aspects of Quantum Chaos)
HPRN-CT-2000-00103.}
\thanks{Research of S.O'K. supported by EPSRC Research Grant GR/R67279/01, and the
EC Research Training Network (Mathematical Aspects of Quantum Chaos)
HPRN-CT-2000-00103.}
\thanks{Research of S.Z.  partially supported by  NSF grant \#DMS-0302518.}

\maketitle

\centerline{{\sc\lowercase{with an appendix}}}\vskip10pt
\centerline{{\sc\uppercase{\bf Converse quantum ergodicity}}} \vskip10pt
\centerline{{\sc\lowercase{by Steve Zelditch}}}\vskip10pt

\centerline{\footnotesize April 29, 2004/\today}

\section*{Abstract}
For a general class of unitary quantum maps, whose underlying
classical phase space is divided into several invariant domains of positive measure, 
we establish analogues of Weyl's law for the distribution
of eigenphases. If the map has one ergodic component, and is periodic
on the remaining domains, we prove the Schnirelman-Zelditch-Colin de Verdi\`ere
Theorem on the equidistribution of eigenfunctions with respect to
the ergodic component of the classical map (quantum ergodicity).
We apply our main theorems to quantised linked twist maps on the
torus. In the Appendix, S. Zelditch connects these studies to some earlier results on `pimpled
spheres' in the setting of Riemannian manifolds. The common
feature is a divided phase space with a periodic component. 

\section{Introduction}

\begin{figure}
\begin{center}
\begin{minipage}{0.49\textwidth}
\unitlength0.1\textwidth
\begin{picture}(10,10)(0,0)
\put(0.5,0.5){\includegraphics[width=0.9\textwidth]{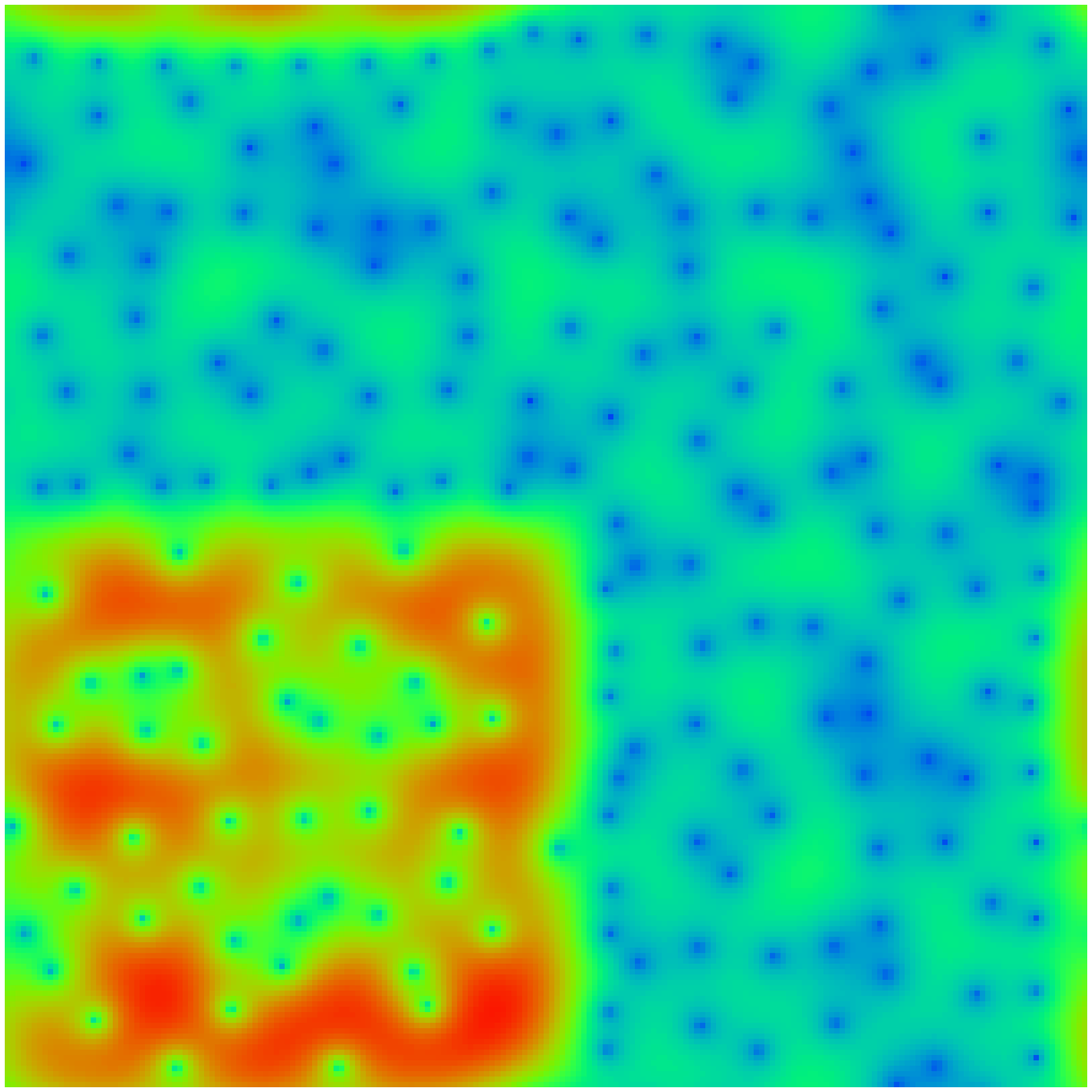}}
\put(0,9){$\uparrow$} \put(0,8.5){$q$}
\put(8.5,0){$p\rightarrow$}
\end{picture}
\end{minipage}
\begin{minipage}{0.49\textwidth}
\unitlength0.1\textwidth
\begin{picture}(10,10)(0,0)
\put(0.5,0.5){\includegraphics[width=0.9\textwidth]{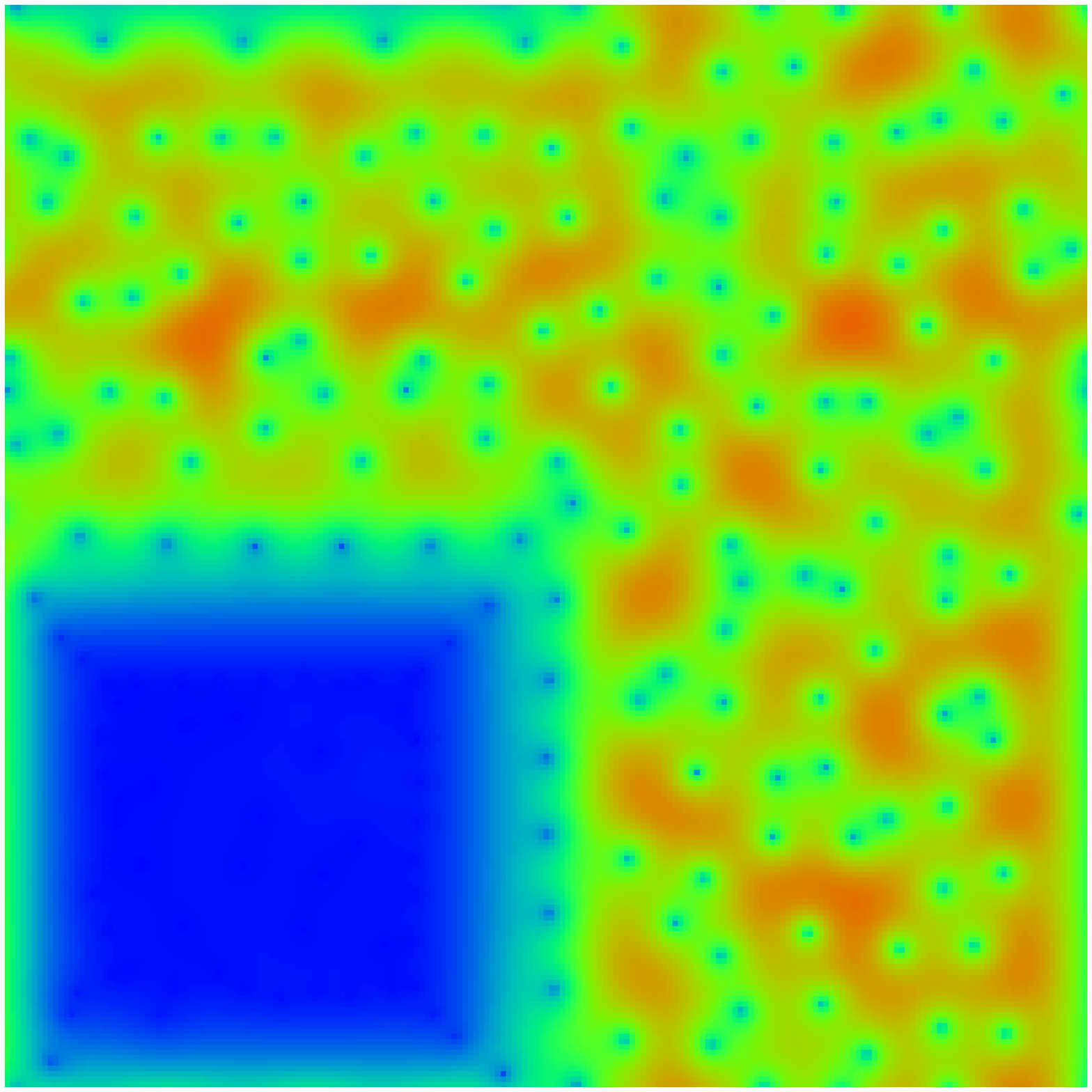}}
\put(0,9){$\uparrow$} \put(0,8.5){$q$}
\put(8.5,0){$p\rightarrow$}
\end{picture}
\end{minipage}
\end{center}
\caption{Two typical quantum eigenstates of a linked twist map on the torus
for $N=201$,
one localised in the lower left quadrant $\scrD_1$ (left image), the other
in the complement $\scrD_0$ (right image). According to Theorem \ref{locthm},
approximately 25\% resp.~ 75\% of all eigenstates behave in this way---the
latter states are in fact equidistributed on $\scrD_0$,
cf.~Theorem \ref{qethm}. \label{figQE}}
\end{figure}

\begin{figure}
\begin{center}
\begin{minipage}{0.49\textwidth}
\unitlength0.1\textwidth
\begin{picture}(10,10)(0,0)
\put(0.5,0.5){\includegraphics[width=0.9\textwidth]{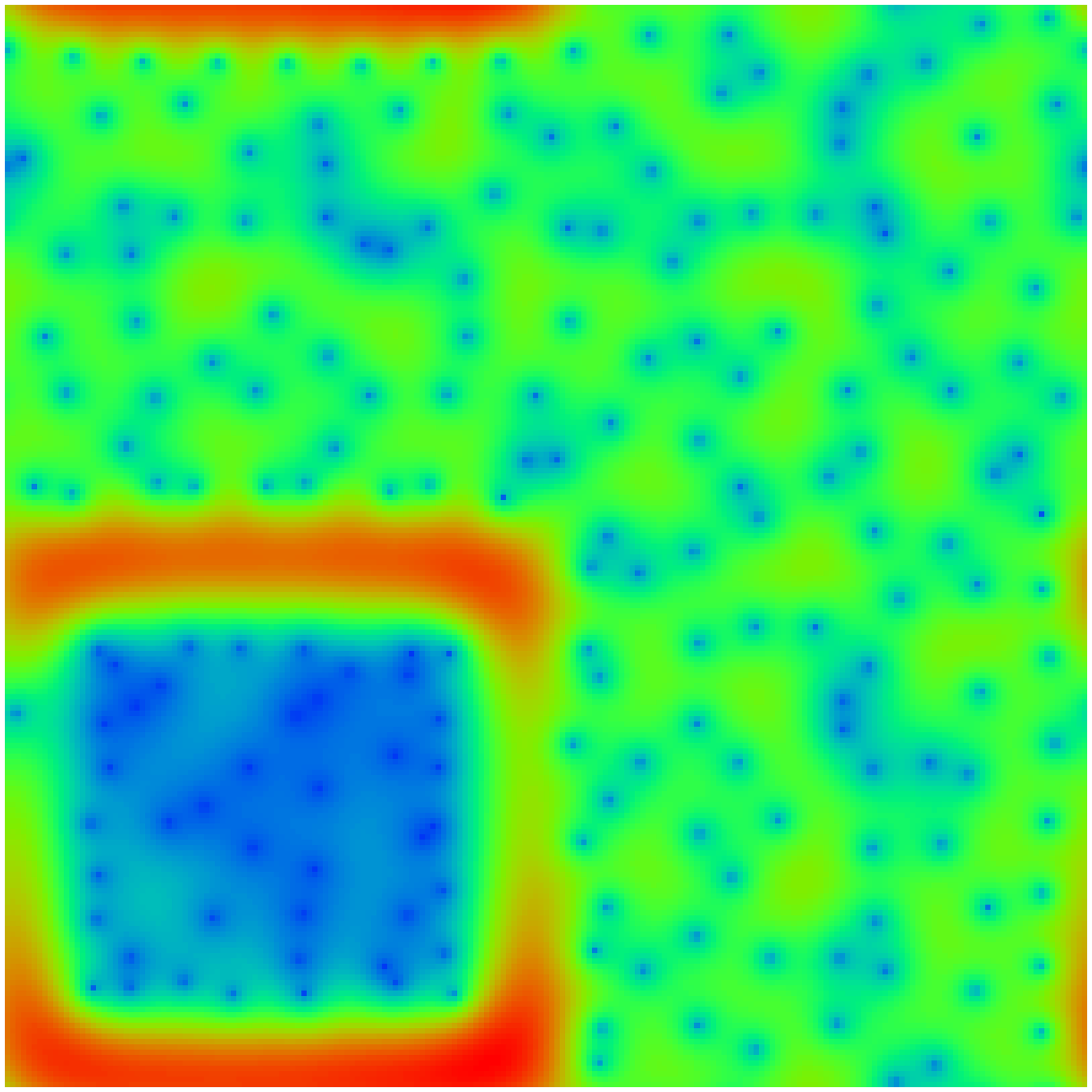}}
\put(0,9){$\uparrow$} \put(0,8.5){$q$}
\put(8.5,0){$p\rightarrow$}
\end{picture}
\end{minipage}
\begin{minipage}{0.49\textwidth}
\unitlength0.1\textwidth
\begin{picture}(10,10)(0,0)
\put(0.5,0.5){\includegraphics[width=0.9\textwidth]{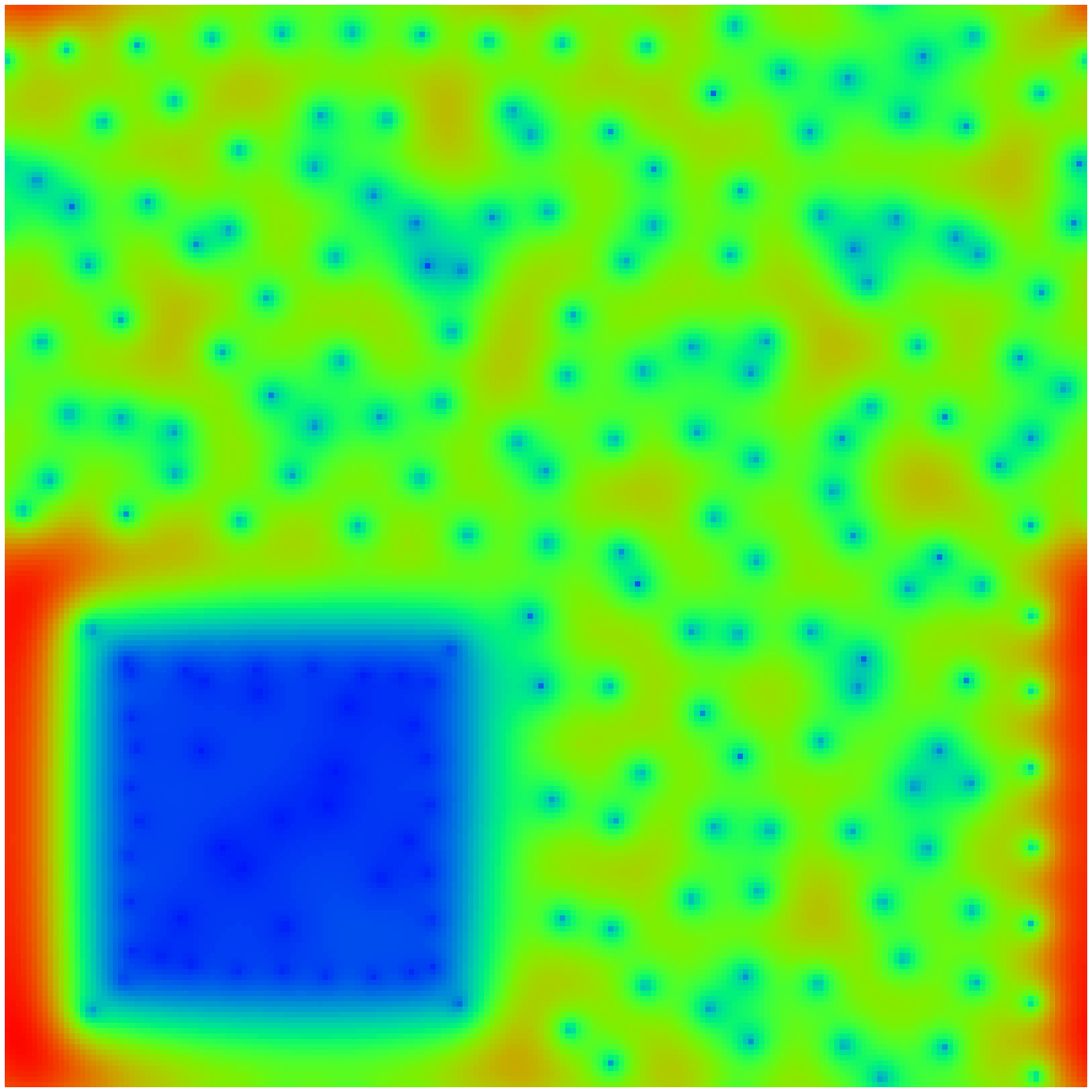}}
\put(0,9){$\uparrow$} \put(0,8.5){$q$}
\put(8.5,0){$p\rightarrow$}
\end{picture}
\end{minipage}
\end{center}
\caption{Two untypical quantum eigenstates of the same linked twist map as
in Figure \ref{figQE}, with $N=201$.
These eigenstates are localised near the boundary between the domains
$\scrD_0$ and $\scrD_1$; the eigenstate on the right has comparable
mass in both domains.
Eigenstates of this type form a sequence of density zero,
cf.~Theorem \ref{locthm}. \label{figQE2}}
\end{figure}

\begin{figure}
\begin{center}
\begin{minipage}{0.8\textwidth}
\unitlength0.1\textwidth
\begin{picture}(10,7.5)(0,0)
\put(0,7.3){\includegraphics[angle=270,width=\textwidth]{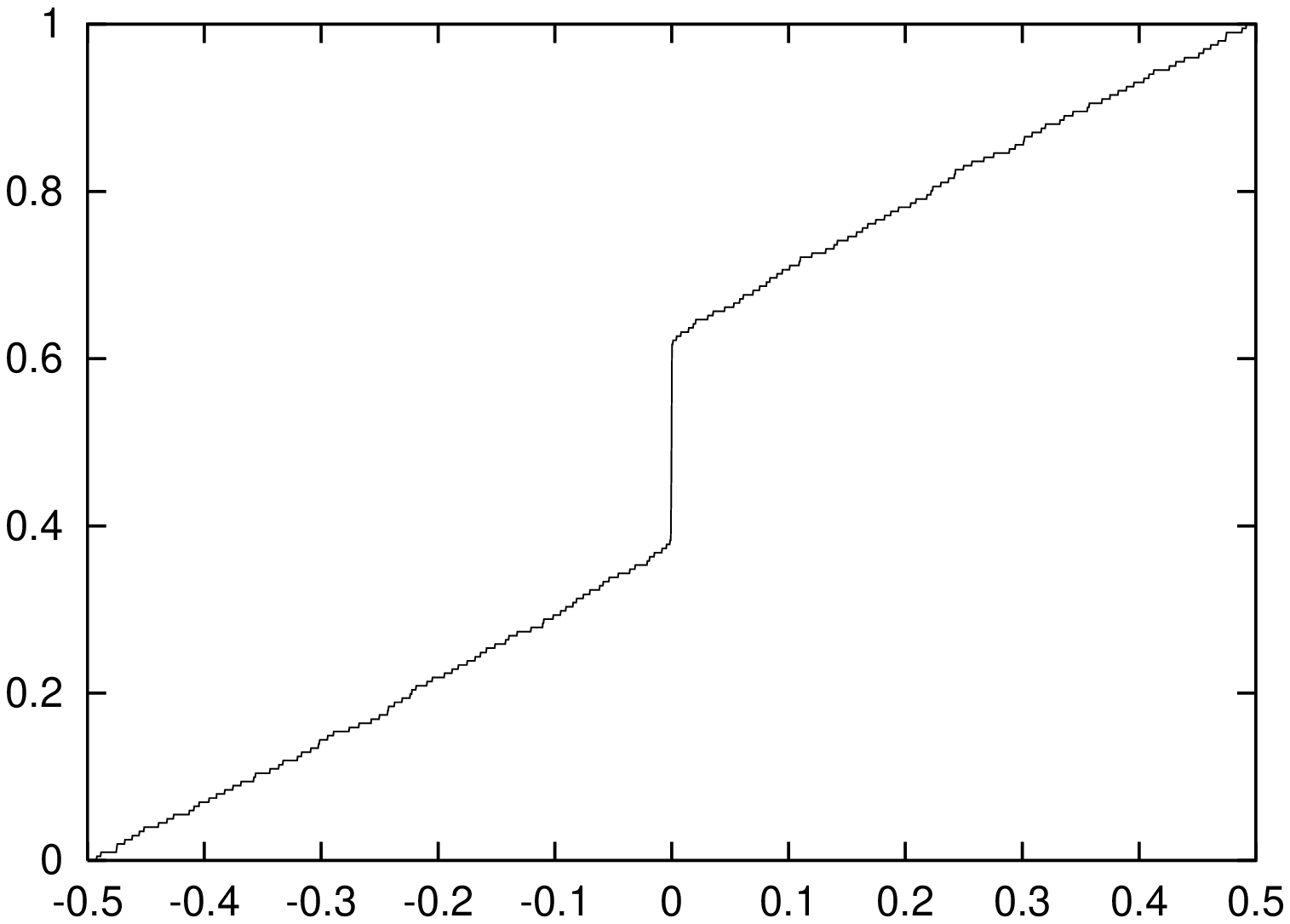}}
\put(0,6.3){$\scrI_N(\theta)$}
\put(9,0){$\theta$}
\end{picture}
\end{minipage}
\end{center}
\caption{The integrated density of quantum eigenphases $\scrI_N(\theta)$
of the linked twist map in Figure \ref{figQE}, with $N=201$.
The jump at $\theta=0$ by $\tfrac14$ reflects the fact that
25\% of eigenphases are asymptotically zero.
The remaining 75\% of eigenphases are uniformly distributed (modulo one) in
the interval $[-\tfrac12,\tfrac12]$, in accordance with Weyl's law
\eqref{dense2}, cf.~Theorem \ref{weyls}. \label{figWeyl}}
\end{figure}

The quantization of an invertible map $\Phi$ on a
$d$-dimensional compact manifold
$\scrM$ produces a unitary $N\times N$ matrix, the {\em quantum map}
$U_N(\Phi)$. If there is a quantization
recipe that works for an infinite sequence of integers $N$,
one natural question is whether dynamical properties of the {\em classical
map} $\Phi$ are recovered in the {\em semiclassical limit} $N\to\infty$.
In this paper we prove that this is indeed possible
for maps whose phase space $\scrM$ is divided into several
$\Phi$-invariant sets of positive measure, provided
the quantization recipe satisfies a {\em correspondence principle}:
quantum and classical evolution of observables must be equivalent
in the semiclassical limit. The precise formulation of the necessary
axioms is given in Section \ref{secSetup}. These are kept fairly general
and allow applications outside of quantum mechanics. In
particular, $\scrM$ is not required to be symplectic.

Let $\varphi_j\in\CC^N$ $(j=1,\ldots,N)$ be an orthonormal basis
of eigenvectors of $U_N(\Phi)$, and $\theta_j\in\RR$
the corresponding eigenphases defined by the relation
\begin{equation}
U_N(\Phi)\, \varphi_j = \e^{2\pi\i \theta_j} \,\varphi_j .
\end{equation}
In Section \ref{secTra} we prove an analog of Weyl's law
\cite{Weyl11,Weyl12,Baltes76,Guillemin85}
for the limiting distribution of the eigenphases as $N\to\infty$.
Let us denote by $\mu$ an invariant probability measure of $\Phi$ on $\scrM$
which is absolutely continuous with respect to Lebesgue measure.
We furthermore assume that
the map is periodic on a collection of sets $\scrD_1,\scrD_2,\ldots$
with positive measure $\mu(\scrD_\nu)>0$ and boundary whose fractal
$d$-dimensional
Minkowski content is zero\footnote{Fractal Minkowski contents and
dimensions also play an important
role in the Weyl-Berry conjecture
for the distribution of eigenvalues of the Laplacian of domains
in $\RR^d$ with fractal boundary, see \cite{Berry79,Berry80,
Lapidus91,Molchanov97}
and references therein.},
and that on the remaining set $\scrD_0$
the periodic orbits of $\Phi$ of any given period
form a set with $d$-dimensional Minkowski content equal to zero.
The number of eigenphases in some generic\footnote{{\em Generic} means
here that $a,b$ are not in the singular support of the limit
density (which in the present case is a countable set).}
interval $[a,b]$ with $b-a\leq 1$ is then
\begin{equation}\label{dense}
\lim_{N\to\infty} \frac{1}{N}
\# \{ j=1,\ldots,N : \theta_j\in[a,b]\bmod 1 \}
= \sum_{\nu=0}^\infty \mu(\scrD_\nu)
\int_a^b \rho_\nu(\theta)\, d\theta
\end{equation}
where $\rho_0(\theta) = 1$ is the uniform probability density mod 1
and, for $\nu\geq 1$, $\rho_\nu(\theta)$ is the uniform probability density
supported on the points $(k/n_\nu)+\alpha_{\scrD_\nu}$; here $n_\nu$ denotes
the period of $\Phi$ on $\scrD_\nu$,
and the constant $\alpha_{\scrD_\nu}\in\RR$ depends on the
chosen quantization recipe. This means in particular that if $\scrM=\scrD_0$,
the spectrum of the quantum map is uniformly distributed on the unit circle.
A formula analogous to \eqref{dense} has been obtained by Zelditch \cite{Zelditch92}, Theorem 3.20,
in the case of the wave group for a compact Riemannian manifold. A special case of formula
\eqref{dense} is proved in \cite{Zelditch97} for quantised contact transformations,
and in \cite{Bouzouina98} for perturbed cat maps; in both cases the set of periodic orbits has
measure zero and thus $\scrM=\scrD_0$.

The main result of this paper describes the semiclassical
distribution of the quantum map's eigenstates on the classical manifold.
If the quantization constants $\alpha_{\scrD_1},\alpha_{\scrD_2},\ldots$
are linearly independent over the rationals, then, for $N$ large,
approximately $N \times \mu(\scrD_\nu)$ of the $N$ eigenstates
localise on the set $\scrD_\nu$ (Section \ref{secLoc}).
If in addition the classical map
acts ergodically on $\scrD_0$, then almost all of the eigenstates
localised on $\scrD_0$ are in fact equidistributed on $\scrD_0$
(Section \ref{secQua}).
This latter result may be viewed as an extension of the
{\em Schnirelman-Zelditch-Colin de Verdi\`ere Theorem},
originally formulated for completely ergodic Hamiltonian flows
\cite{Schnirelman74,Zelditch87,Colin85,Helffer87,Gerard93,Zelditch96b}
and maps
\cite{Bouzouina96,Zelditch96,Zelditch97,DeBievre98}, to maps with
partially ergodic phase space. The possibility that eigenfunctions
localise exclusively on the ergodic or on the integrable component had
been conjectured by Percival in the 1970s \cite{Percival73},
and is known as the {\em semiclassical wave function hypothesis}.

Two typical and two untypical examples
of eigenstates of a quantised linked twist map on a
two-dimensional torus (see Section \ref{appLinked}),
with $N=201$, are displayed
as a Husimi density plot\footnote{See e.g.~\cite{Nonnenmacher98,Baecker03}
for a detailed discussion of Husimi functions.} in Figures \ref{figQE}
and \ref{figQE2}.
The classical map acts as the identity on the lower left quadrant,
and ergodically on the L-shaped complement (Section \ref{appErgodic}).
Figure \ref{figWeyl} shows the integrated density
\begin{equation}
\scrI_N(\theta):=\frac1N
\# \{ j=1,\ldots,N : \theta_j\in[-\tfrac12,\theta]\bmod 1 \}
\end{equation}
of quantum eigenphases of the map
with $\theta$ ranging from $-\tfrac12$ to $\tfrac12$.
In this case, the limit \eqref{dense} evaluates to
\begin{equation}\label{dense2}
\lim_{N\to\infty} \scrI_N(\theta)
=
\begin{cases}
\tfrac34 \theta + \tfrac38& \text{for $\theta\in[-\tfrac12,0)$,}\\
\tfrac34 \theta + \tfrac58 & \text{for $\theta\in(0,\tfrac12]$.}
\end{cases}
\end{equation}
Very similar
quantum maps have been investigated numerically in \cite{Malovrh02};
our theorems apply to those cases with sharply divided phase space
and rational frequencies in all elliptic islands.
A further interesting family of toral maps with
mixed dynamics are the {\em lazy baker maps} \cite{Lak93}.
Here the phase space is divided into countably many rational elliptic
islands of total measure one; the hyperbolic dynamics takes place
on a fractal set of Lebesgue measure zero.

In the case of Hamiltonian flows with partially ergodic phase
space\footnote{Examples of such systems are the billiard flows
in Bunimovich's mushrooms \cite{Bunimovich91}.},
Schubert \cite{Schubert01} has proved a result analogous to our
quantum ergodicity theorem (Theorem \ref{qethm}), which
however only holds for quasimode solutions of the Schr\"odinger equation
and not necessarily for the eigenstates itself. The problem is that
near-degeneracies in the spectrum of the quantum Hamiltonian (which
in general cannot be ruled out) may create eigenstates
that are extended across the entire phase space although the
corresponding quasimodes are localised
on the flow invariant components. For the quantum maps considered here, the
spectrum can be controlled sufficiently well to rule out this possibility.
A similar observation results from the analysis of the wave group on Riemannian manifolds 
where geodesic flow is periodic on some open invariant component, 
see the Appendix by S. Zelditch.

It should be emphasised that our results do not address the question
of the possible localization of eigenstates on sets of measure
zero (such as scarred eigenstates, bouncing ball modes
or the recently discovered hierarchical states \cite{Ketzmerick00}).
Results in this direction have recently been obtained
in the case of cat maps \cite{Bonechi03,Faure03,Faure04} and
piecewise affine maps on the torus \cite{Chang04}, which neatly
complement the proofs of quantum unique ergodicity for cat maps
\cite{Degli95,Kurlberg00,Kurlberg01}, parabolic maps \cite{que}
and the modular surface \cite{Watson02,Lindenstrauss03}
(see the survey \cite{Sarnak03}).

Concrete examples of maps satisfying the axioms set out in Section
\ref{secSetup} are toral linked twist maps, see
Sections \ref{appQuantum} and \ref{appLinked}. The specific
structure of these maps permits a more detailed asymptotic analysis
which will be presented elsewhere \cite{OKeefe}.

The axioms in Section
\ref{secSetup} are in fact sufficiently general to allow also applications to
sequences of unitary matrices without quantum mechanical interpretation.
In Section \ref{appDis} we discuss an application arising
in the discretization of classical evolution operators.

\section*{Acknowledgements}

We thank A. B\"acker, M. van den Berg, E. Bogomolny, Y. Colin de Verdi\`ere,
S. De Bi\`evre, F. Mezzadri, S. Nonnenmacher, J. Robbins, Z. Rudnick, R. Schubert,
F. Vivaldi, A. Voros and S. Zelditch for stimulating discussions.

\section{Set-up}\label{secSetup}

Let $\scrM$ be a $d$-dimensional
compact smooth manifold, and $\mu$ a probability measure
on $\scrM$ which is absolutely continuous with respect to Lebesgue
measure.

We fix an atlas of local charts $\phi_j:\scrV_j \to \RR^d$,
where the open subsets $\scrV_j$ cover $\scrM$.
In the following we thus identify subsets $\scrS$ of $\scrM$ with subsets
$\Sigma$ of $\RR^d$ in the standard way.
Let $\Sigma$ be a subset of $\RR^d$, and
\begin{equation}
\Sigma(\epsilon)=\{ \xi \in\RR^d : d(\xi,\Sigma) \leq \epsilon \}
\end{equation}
its closed $\epsilon$-neighbourhood,
where $d(\,\cdot\, , \,\cdot\,)$ is the euclidean metric on $\RR^d$.
The {\em $s$-dimen\-sional upper Minkowski content} of $\Sigma$ is defined as
\begin{equation}
\fraM^{* s}(\Sigma) := \limsup_{\epsilon\to 0}
(2\epsilon)^{s-d}\nu\big(\Sigma(\epsilon)\big) ,
\end{equation}
where $\nu$ is Lebesgue measure,
see \cite[Section 5.5]{Mattila95}.
We say $\Sigma$
{\em has Minkowski content zero} if $\fraM^{* d}(\Sigma)=0$.
This is equivalent to saying that for every $\delta>0$ we
can cover $\Sigma$ with equi-radial euclidean balls of
total measure less than $\delta$.
We say {\em a subset $\scrS$ of $\scrM$ has Minkowski content zero}
if each of the sets $\Sigma_j:=\phi_j(\scrS|_{\scrV_j})\subset\RR^d$
has Minkowski content zero.

We consider piecewise smooth invertible maps
$\Phi: \scrM \to \scrM$ which preserve $\mu$.
By {\em piecewise smooth} we mean here and in the following
that  there is a partitioning of $\scrM$ into countably many open sets
$\scrU_i$, i.e.,
$\scrM = \overline{\bigcup_i \scrU_i}$ and $\scrU_i \cap \scrU_j = \emptyset$,
so that $\Phi|_{\overline \scrU_i}$ is smooth,%
\footnote{This means that
$\Phi|_{\overline \scrU_i}$ and all its derivatives are
bounded continuous functions $\overline \scrU_i \to\scrM$;
we allow for the possibility that those bounds are not uniform in $i$.}
and the boundary set
$\overline{\bigcup_i \partial\scrU_i}$
has Minkowski content zero. We will refer to this set as
the {\em domain of discontinuity of $\Phi$}, and call its complement
$\scrM-\overline{\bigcup_i \partial\scrU_i}$ the
{\em domain of continuity of $\Phi$}.

Let $\M_N(\CC)$ be the space of
$N\times N$ matrices with complex coefficients.
For a given infinite subset ({\em index set}) $\scrI\subset\NN$,
we say two sequences of matrices,
\begin{equation}
\vecA := \{ A_N \}_{N\in\scrI}, \qquad \vecB:=\{ B_N \}_{N\in\scrI},
\end{equation}
are {\em semiclassically equivalent}, if
\begin{equation}
\| A_N - B_N \| \to 0
\end{equation}
as $N\in\scrI$ tends to infinity,
where $\|\,\cdot\,\|$ denotes the usual operator norm
\begin{equation}
\|A\|:= \sup_{\psi\in\CC^N-\{0\}} \frac{\| A \psi \|}{\|\psi\|}.
\end{equation}
We denote this equivalence relation by
\begin{equation}
\vecA \sim \vecB .
\end{equation}

\begin{lem}
If $\vecA \sim \vecB$ then $\Tr A_N = \Tr B_N + o(N)$.
\end{lem}

\begin{proof}
We have
\begin{equation}
\frac1N |\Tr A_N - \Tr B_N| \leq  \| A_N-B_N \| \to 0.
\end{equation}
\end{proof}

Let us define the product of two matrix sequences by
$\vecA\vecB=\{ A_N B_N \}_{N\in\scrI}$, the inverse of $\vecA$ by
$\vecA^{-1}=\{ A_N^{-1} \}_{N\in\scrI}$, and its hermitian
conjugate by $\vecA^\dagger=\{ A_N^\dagger \}_{N\in\scrI}$.

\begin{axiom}[The correspondence principle for quantum observables]
\label{quantass}
Fix a measure $\mu$ as above.
For some index set $\scrI\subset\NN$,
there is a sequence $\vecOp:=\{ \Op_N \}_{N\in\scrI}$ of linear maps,
\begin{equation*}
\Op_N: \C^\infty(\scrM) \to \M_N(\CC),
\qquad
a \mapsto \Op_N(a) ,
\end{equation*}
so that
\begin{itemize}
\item[(a)]
for all $a\in\C^\infty(\scrM)$,
\begin{equation*}
\vecOp(\overline a) \sim \vecOp(a)^\dagger;
\end{equation*}
\item[(b)]
for all $a_1,a_2\in\C^\infty(\scrM)$,
\begin{equation*}
\vecOp(a_1) \vecOp(a_2) \sim \vecOp(a_1 a_2) ;
\end{equation*}
\item[(c)]
for all $a\in\C^\infty(\scrM)$,
\begin{equation*}
\lim_{N\to\infty} \frac1N \Tr \Op_N(a) = \int_\scrM a\, d\mu .
\end{equation*}
\end{itemize}
\end{axiom}

Examples of quantum observables satisfying these conditions are
given in Section \ref{appQuantum}. In standard quantization recipes
(such as the one discussed in Section \ref{appQuantum})
one in addition has the property that
\begin{equation}
\vecOp(a_1)\vecOp(a_2)-\vecOp(a_2)\vecOp(a_1) \sim
\frac{1}{2\pi\i N}\,\vecOp(\{a_1, a_2\})
\end{equation}
where $\{\;,\;\}$ is the Poisson bracket.
This assumption is however not necessary for any of the results
proved in this paper. The axioms (a)--(c) in fact
apply to examples without quantum mechanical significance.
One interesting case arises in the discretization
of linked twist maps, where
\begin{equation}
\vecOp(a_1)\vecOp(a_2)= \vecOp(a_1 a_2)=\vecOp(a_2)\vecOp(a_1),
\end{equation}
see Section \ref{appDis}.

\begin{axiom}[The correspondence principle for quantum maps] \label{corresp}
There is a sequence of unitary matrices
$\vecU(\Phi):=\{ U_N(\Phi) \}_{N\in\scrI}$
such that for any $a\in\C^\infty(\scrM)$ with compact support contained in
the domain of continuity of $\Phi$, we have
\begin{equation*}
\vecU(\Phi)^{-1} \vecOp(a) \vecU(\Phi) \sim \vecOp(a\circ \Phi) .
\end{equation*}
\end{axiom}

In the following we consider maps $\Phi$ that may be
periodic on a collection of
disjoint sets $\scrD_\nu\subset\scrM$ ($\nu=1,2,\ldots$)
of positive measure $\mu(\scrD_\mu)>0$, with periods $n_\nu$
so that $\Phi^{n_\nu}\big|_{\scrD_\nu}=\id$.
In addition to Axioms \ref{quantass} and \ref{corresp},
we will here stipulate that
there are constants $\alpha_{\scrD_\nu}\in\RR$ such that
\begin{equation} \label{spezi}
\vecU(\Phi)^{n_\nu} \vecOp(a) \sim e(n_\nu \alpha_{\scrD_\nu})\, \vecOp(a),
\qquad e(z):=\exp(2\pi\i z),
\end{equation}
for any $a\in\C^\infty(\scrM)$ with compact support
contained in $\scrD_\nu$ and
the domains of continuity of $\Phi,\Phi^2,\ldots,\Phi^{n_\nu}$.
Whereas Axioms \ref{quantass} and \ref{corresp}
are satisfied by all standard quantization schemes, condition \eqref{spezi} is
more restrictive: the constant $\alpha_{\scrD_\nu}$
could for instance be replaced by $\vecOp(\beta_\nu)$ where $\beta_\nu$ is a
non-constant smooth function on $\scrD_\nu$; note the relation
\begin{equation} \label{spezi2}
\vecU(\Phi)^{n_\nu} \vecOp(a) \sim e(n_\nu \vecOp(\beta_\nu))\,
\vecOp(a)
\end{equation}
is still consistent with Axiom \ref{corresp}. Condition \eqref{spezi}
is however essential in the proofs of our main results since the spectrum
of $e(\vecOp(\beta_\nu))$ may, in general, be dense on the unit circle
in $\CC$.

In Sections \ref{appLinked} and \ref{appDis} we discuss examples
of semiclassical sequences of quantum maps satisfying the above
Axioms \ref{quantass}, \ref{corresp} and condition \eqref{spezi}.

\section{Example: Linked twist maps}\label{secExample}

In this section we construct a well known example of quantum observables
on the two-dimensional torus $\scrM=\TT^2:=\RR^2/\ZZ^2$
satisfying Axiom \ref{quantass} (cf.~\cite{Degli03}),
and corresponding examples of quantum linked twist maps satisfying
Axiom \ref{corresp}.

\subsection{Quantum tori}\label{appQuantum}
It is convenient to represent a vector $\psi\in\CC^N$ as a function
$\psi: \ZZ/N\ZZ \to \CC$.
Let us define the translation operators
\begin{equation}
[t_1 \psi](Q)=\psi(Q+1)
\end{equation}
and
\begin{equation}
[t_2\psi](Q)=e_N(Q)\psi(Q),
\end{equation}
where $e_N(x):=e(x/N)=\exp(2\pi\i x/N)$.
One easily checks that
\begin{equation}
\label{heisenberg}
t_1^{m_1}t_2^{m_2}=t_2^{m_2}t_1^{m_1}e_N(m_1 m_2)\qquad \forall m_1,m_2\in\ZZ.
\end{equation}
These relations are known as the {\em Weyl-Heisenberg commutation relations}.
For $\vecm =(m_1,m_2)\in \ZZ^2$ put
\begin{equation}
T_N(\vecm)=e_N\left(\frac{m_1 m_2}{2}\right) t_2^{m_2} t_1^{m_1} .
\end{equation}
Then
\begin{equation}
T_N(\vecm)T_N(\vecn)=e_N\left(\frac{\omega(\vecm,\vecn)}{2}\right)
T_N(\vecm+\vecn)
\end{equation}
with the symplectic form
\begin{equation}
\omega(\vecm, \vecn)=m_1 n_2-m_2 n_1 .
\end{equation}
For any  $a\in\C^\infty(\TT^2)$, we define the quantum observable
\begin{equation}
\Op_N(a)=\sum_{\vecm\in\ZZ^2} \widehat{a}(\vecm)T_N(\vecm)
\end{equation}
where
\begin{equation}
\widehat{a}(\vecm) = \int_{\TT^2} a(\vecxi) e(-\vecxi\cdot\vecm) \, d\xi
\end{equation}
are the Fourier coefficients of $a$. The
observable $\Op_N(a)$ is also called the \textit{Weyl quantization of} $a$.
Axiom \ref{quantass} (a) is trivially satisfied.
Axioms \ref{quantass} (b) and (c) follow from the following lemmas.

\begin{lem}
For all $a_1,a_2\in\C^\infty(\TT^2)$
\begin{equation}
\| \Op_N(a_1)\Op_N(a_2) - \Op_N(a_1 a_2) \|
\leq \frac{\pi}{N}
\bigg( \sum_{\vecm\in\ZZ^2} \|\vecm\| |\widehat a_1(\vecm)| \bigg)
\bigg( \sum_{\vecn\in\ZZ^2} \|\vecn\| |\widehat a_2(\vecn)| \bigg) .
\end{equation}
\end{lem}

\begin{proof}
Using the commutation relations (\ref{heisenberg}) we find
\begin{align}
\Op_N(a_1)\Op_N(a_2)
& = \sum_{\vecm,\vecn\in\ZZ^2}
\widehat{a_1}(\vecm)\widehat{a_2}(\vecn) T_N(\vecm) T_N(\vecn) \\
& = \sum_{\vecm,\vecn\in\ZZ^2}
e_N\left(\frac{\omega(\vecm,\vecn)}{2}\right)
\widehat{a_1}(\vecm)\widehat{a_2}(\vecn) T_N(\vecm+\vecn)\\
& =\sum_{\vecm,\veck \in \ZZ^2}e_N\left(\frac{\omega(\vecm,\veck)}{2}\right)
\widehat{a_1}(\vecm)\widehat{a_2}(\veck-\vecm) T_N(\veck)
\end{align}
with $\veck=\vecn+\vecm$, and hence
\begin{equation}
\| \Op_N(a_1)\Op_N(a_2) - \Op_N(a_1 a_2) \|
\leq
\sum_{\vecm,\vecn\in\ZZ^2}
\bigg|e_N\left(\frac{\omega(\vecm,\vecn)}{2}\right) -1 \bigg|\;
\big| \widehat{a_1}(\vecm)\big| \; \big|\widehat{a_2}(\vecn) \big|
\end{equation}
The lemma now follows from
\begin{equation}
|e(x)-1|\leq |2\pi x|, \qquad
|\omega(\vecm,\vecn)| \leq \|\vecm\|\,\|\vecn\| .
\end{equation}
\end{proof}

\begin{lem}
For any $a\in\C^\infty(\TT^2)$ and $R>1$
\begin{equation}
\frac1N \Tr \Op_N(a) = \int_{\TT^2} a\, d\mu + O_{a,R}(N^{-R}).
\end{equation}
\end{lem}

\begin{proof}
Note that
\begin{equation}
\Tr T_N(\vecm)=
\begin{cases}
N & \text{ if $\vecm=\vecnull\bmod N\ZZ^2$,}\\
0 & \text{ otherwise.}
\end{cases}
\end{equation}
The lemma now follows from the rapid decay of the Fourier coefficients
$\widehat a(\vecm)$ for $\|\vecm\|\to\infty$.
\end{proof}

Note that we have the alternative representation for $\Op_N(a)$,
\begin{equation}\label{alter}
[\Op_N(a)\psi](Q) = \sum_{m\in\ZZ}
\widetilde a\bigg(m,\frac{Q}{N}+\frac{m}{2N}\bigg) \psi(Q+m)
\end{equation}
where
\begin{equation}
\widetilde a(m,q)= \int_{\RR/\ZZ} a(p,q) \, e(-pm)\, dp ,
\end{equation}
which is sometimes useful. In fact \eqref{alter} permits
to quantise observables $a$ which are discontinuous
in the $q$-variable.
Note that if $a$ is a smooth function of $p$ and,
for any $\nu\geq 0$, $\frac{d^\nu}{dp^\nu}a(p,q)$
is a bounded function on $\TT^2$,
then, for any $R>1$, there is a constant $C_R$ such that
\begin{equation}\label{rdec}
|\widetilde a(m,q)|\leq C_R (1+|m|)^{-R}
\end{equation}
for all $m,q$.
This fact is proved using integration by parts.
Of course \eqref{rdec} holds in particular
for smooth observables $a\in\C^\infty(\TT^2)$.

\subsection{Quantum linked twist maps}\label{appLinked}

A {\em twist map} $\Psi_f$ is a map $\TT^2\to \TT^2$ defined by
\begin{equation}
\Psi_f : \begin{pmatrix} p \\ q \end{pmatrix}
\mapsto \begin{pmatrix} p+ f(q) \\ q  \end{pmatrix} \mod \ZZ^2
\end{equation}
where $f:\RR/\ZZ\to \RR$ is piecewise smooth, i.e., the
domain of discontinuity of $f$ in
$\RR/\ZZ$ has Minkowski content zero, cf.~Section \ref{secSetup}.
Piecewise smooth functions of this type may be realised by
taking a countable set of points $0=\xi_0 < \xi_1 < \ldots < \xi_\infty=1$
with finitely many accumulation points in $[0,1]$, and assume
that $f\in\C^\infty([\xi_i,\xi_{i+1}])$ for all $i=0,1,\ldots,\infty$.

Obviously Lebesgue measure $d\mu=dp\,dq$ is invariant under $\Psi_f$.
A {\em linked twist map} $\Phi$ is now obtained by combining two twist
maps, $\Psi_{f_1}$ and $\Psi_{f_2}$, by setting
\begin{equation}\label{ltm}
\Phi = \R \circ \Psi_{f_1} \circ \R^{-1} \circ  \Psi_{f_2}
\end{equation}
with the rotation
\begin{equation} \label{rott}
\R: \begin{pmatrix} p \\ q  \end{pmatrix}
\mapsto \begin{pmatrix} q \\ -p \end{pmatrix} \mod \ZZ^2 .
\end{equation}
Since $\Psi_{f_1}$, $\Psi_{f_2}$ and $\R$ preserve $\mu$, so does $\Phi$.
More explicitly, we have
\begin{equation}
\R \circ \Psi_{f} \circ \R^{-1}
: \begin{pmatrix} p \\ q  \end{pmatrix}
\mapsto \begin{pmatrix} p \\ q-f(p) \end{pmatrix} \mod \ZZ^2
\end{equation}
and thus
\begin{equation}
\Phi: \begin{pmatrix} p \\ q  \end{pmatrix}
\mapsto \begin{pmatrix} p+f_2(q) \\ q-f_1\big(p+f_2(q)\big)
\end{pmatrix} \mod \ZZ^2 .
\end{equation}

We define the quantization of the twist map $\Psi_f$ by
the unitary operator
\begin{equation}
[U_N(\Psi_f)]\psi(Q)= e\left[-N V\left(\frac{Q}{N}\right) \right]\psi(Q)
\end{equation}
where $V$ is an arbitrary choice of a piecewise smooth function
$\RR/\ZZ\to\RR$ satisfying $f=-V'$ and $V\in\C^\infty$
on the domain of continuity of $f$.%
\footnote{In the case of smooth twist maps $\Psi_f:\TT^2\to \TT^2$
it is more convenient to view $f$ as a $\C^\infty$ function
$\RR/\ZZ\to\RR/\ZZ$, thus avoiding the introduction of artificial
discontinuities in $f$. The potential $V$ may now be defined
as a $\C^\infty$ function $V:\RR\to\RR$ with $f=-V'$ locally,
and the condition that
$N V(\frac{Q}{N}+1)=N V(\frac{Q}{N})\bmod 1$
for every $Q,N$. Examples are $f(q)=2\tau q$ with $\tau\in\ZZ$,
for which $V(q)=-\tau q^2$. In this case the correspondence principle
stated in Proposition \ref{ET} is
in fact exact, i.e., the right hand side of \eqref{ETeq} is
identically zero.}

\begin{prop}\label{ET}
For any $a\in\C^{\infty}(\TT^2)$ with compact support contained in
the domain of continuity of $\Psi_f$, we have
\begin{equation}\label{ETeq}
\| U_N(\Psi_f)^{-1} \Op_N(a) U_N(\Psi_f) - \Op_N(a\circ \Psi_f) \|
= O(N^{-2})
\end{equation}
where the implied constant depends on $a$.
\end{prop}

\begin{proof}
We have
\begin{multline}
[U_N(\Psi_f)^{-1} \Op_N(a) U_N(\Psi_f)\psi](Q)\\
= \sum_{m\in\ZZ}
\widetilde a\bigg(m,\frac{Q}{N}+\frac{m}{2N}\bigg)
e\left\{-N \left[ V\left(\frac{Q+m}{N}\right) -V\left(\frac{Q}{N}\right)
\right] \right\}
\psi(Q+m),
\end{multline}
and
\begin{equation}
\Op_N(a\circ \Psi_f) =
 \sum_{m\in\ZZ}
\widetilde a\bigg(m,\frac{Q}{N}+\frac{m}{2N}\bigg)
e\left[m f\left(\frac{Q}{N}+\frac{m}{2N}\right)\right]
\psi(Q+m) ,
\end{equation}
since
\begin{equation}
\widetilde{(a\circ \Psi_f)}(m,q)= e[m f(q)]\,\widetilde a(m,q) .
\end{equation}
Therefore
\begin{equation}
\| U_N(\Psi_f)^{-1} \Op_N(a) U_N(\Psi_f) - \Op_N(a\circ \Psi_f) \|
\leq
\max_q \sum_{m\in\ZZ}
\left| \widetilde a\bigg(m,q+\frac{m}{2N}\bigg)
c_m(q,N) \right|
\end{equation}
with
\begin{equation}
c_m(q,N) =
e\left\{-N \left[ V\left(q+\frac{m}{N}\right) -V\left(q\right)
\right] \right\} - e\left[m f\left(q+\frac{m}{2N}\right)\right] .
\end{equation}
Since $|c_m(q,N)|\leq 2$ and $|\widetilde a(m,q)|\leq (1+|m|)^{-5}$,
we have
\begin{equation}
\max_q \sum_{|m|\geq N^{1/2}}
\left| \widetilde a\bigg(m,q+\frac{m}{2N}\bigg)
c_m(q,N) \right|
\leq
N^{-2} .
\end{equation}
Let us denote by $C\!S$ the projection of the compact support
of $a$ onto the $q$ axis. $C\!S$ is a compact set which is in
the domain of continuity of $f$.

For $|m|<N^{1/2}$, Taylor expansion around $x=q+\frac{m}{2N}$ yields
(the second order terms cancel)
\begin{align}
V\left(x+\frac{m}{2N}\right) -V\left(x-\frac{m}{2N}\right)
& = V'\left(x\right) \frac{m}{N} + O\left(\frac{m^3}{N^3}\right) \\
& = -f\left(x\right) \frac{m}{N} + O\left(\frac{m^3}{N^3}\right) .
\end{align}
uniformly for all $|m|<N^{1/2}$ and all $q\in C\!S$,
provided $N$ is sufficiently large so that
$[q-N^{-1/2},q+N^{-1/2}]$ is contained in the domain of
continuity.
Hence in this case
\begin{equation}
c_m(q,N) = O\left(\frac{m^3}{N^2}\right)
\end{equation}
and
\begin{align}
\max_q \sum_{|m| < N^{1/2}}
\left| \widetilde a\bigg(m,q+\frac{m}{2N}\bigg)
c_m(q,N) \right|
& \leq
O(N^{-2})
\max_q \sum_{m\in\ZZ}
\left| m^3 \widetilde a\bigg(m,q+\frac{m}{2N}\bigg)
\right| \\
& =O(N^{-2}) .
\end{align}
\end{proof}

\begin{prop}\label{ETT}
Suppose $V(q)=v=\const$ for all $q$ in some open interval
$I \subset\RR/\ZZ$. Then,
for any $T>1$ and any
$a\in\C^{\infty}(\TT^2)$ with compact support contained in
$\RR/\ZZ\times I\subset\TT^2$, we have
\begin{equation}
\| U_N(\Psi_f)\Op_N(a) - e(-N v) \Op_N(a) \| = O(N^{-T})
\end{equation}
where the implied constant depends on $a$ and $T$.
\end{prop}

\begin{proof}
We have
\begin{multline}
[(U_N(\Psi_f)- e(-Nv)) \Op_N(a)\psi](Q)\\
= \sum_{m\in\ZZ}
\widetilde a\bigg(m,\frac{Q}{N}+\frac{m}{2N}\bigg)
\left\{ e\left[ -N V\left(\frac{Q+m}{N}\right)\right]-e(-N v) \right\}
\psi(Q+m).
\end{multline}
As before we split the sum into two terms
corresponding to $|m|<N^{1/2}$ and $|m|\geq N^{1/2}$.
For $N$ large enough, the first term vanishes since
$\widetilde a(m,q)$, as a function of $q$,
is compactly supported inside the open interval $I$.
The second term is bounded by
\begin{equation}
\max_q \sum_{|m| \geq N^{1/2}}
\left| \widetilde a\bigg(m,q+\frac{m}{2N}\bigg)
\left\{ e\left[ -N V(q)\right]-e(-N v) \right\}  \right|
\ll_{a,T} N^{-T}
\end{equation}
for any $T>1$ due to the rapid decay \eqref{rdec}.
\end{proof}

The discrete Fourier transform $\scrF_N$ is a unitary operator
defined by
\begin{equation}
[\scrF_N \psi](P)= \frac{1}{\sqrt N} \sum_{Q=0}^{N-1} \psi(Q) e_N(-QP) .
\end{equation}
Its inverse is given by the formula
\begin{equation}
[\scrF_N^{-1} \psi](Q)= \frac{1}{\sqrt N} \sum_{P=0}^{N-1} \psi(P) e_N(PQ) .
\end{equation}

\begin{prop}\label{FT}
For any $a\in\C^\infty(\TT^2)$
\begin{equation}
\scrF_N^{-1} \Op_N(a) \scrF_N = \Op_N(a\circ\R)
\end{equation}
with the rotation $\R$ as in \eqref{rott}.
\end{prop}

\begin{proof}
This follows from the identities $\scrF_N^{-1} t_1 \scrF_N = t_2^{-1}$
and $\scrF_N^{-1} t_2 \scrF_N = t_1$.
\end{proof}

The Fourier transform may therefore be viewed as a quantization of
the rotation $\R$ which satisfies an {\em exact} correspondence
principle, cf.~Axiom \ref{corresp}.

The quantization of the linked twist map is now defined by
\begin{equation}
U_N(\Phi)=\scrF_N\,  U_N(\Psi_{f_1})\, \scrF_N^{-1}\,  U_N(\Psi_{f_2}) .
\end{equation}
\begin{prop}
For any $a\in\C^{\infty}(\TT^2)$ with compact support in the
domain of continuity of $\Phi$, we have
\begin{equation}
\| U_N(\Phi)^{-1} \Op_N(a) U_N(\Phi) - \Op_N(a\circ \Phi) \|
= O(N^{-2})
\end{equation}
where the implied constant depends on $a$.
\end{prop}

\begin{proof}
Apply Propositions \ref{ET} and \ref{FT}.
\end{proof}

The quantum map $U_N(\Phi)$ thus satisfies Axiom \ref{corresp}.

\begin{prop}\label{ETTT}
Suppose $V_1(p)=v_1=\const$ for all $p$ in some open interval
$I_1 \subset\RR/\ZZ$, and $V_2(q)=v_2=\const$ for all $q$ in some open interval
$I_2 \subset\RR/\ZZ$. Then,
for any $T>1$ and any
$a\in\C^{\infty}(\TT^2)$ with compact support contained in the rectangle
$I_1\times I_2 \subset \TT^2$, we have
\begin{equation}
\| U_N(\Phi)\Op_N(a) - e[-N(v_1+v_2)] \Op_N(a) \| = O(N^{-T})
\end{equation}
where the implied constant depends on $a$ and $T$.
\end{prop}

\begin{proof}
Apply Propositions \ref{ETT} and \ref{FT}.
\end{proof}

In the examples considered in Section \ref{appErgodic} we have $v_1=v_2=0$,
and furthermore, for any $n\neq 0$, $\Phi^n$ acts as the identity precisely
on the rectangle $I_1 \times I_2$.
Hence condition \eqref{spezi}
[cf.~condition (c) of our central Theorem \ref{weyls} below]
is satisfied in this case.
In cases where $v_1+v_2\neq 0 \bmod 1$ one may consider subsequences
of $N\to\infty$ for which $\{ N(v_1+v_2) \} \to \alpha$, for any suitable
fixed $\alpha\in[0,1]$; here $\{\,\cdot\,\}$ denotes the fractional part.

\subsection{Ergodic properties of linked twist maps}\label{appErgodic}

The ergodic properties of linked twist maps are well understood
\cite{Burton80,Przytycki83}.
Let $[a_i,b_i]$ ($i=1,2$) be subintervals of $\RR/\ZZ$,
and choose functions $f_i:\RR/\ZZ\to\RR$ with
\begin{itemize}
\item[(a)]
$f_i(q)=0$ for $q\notin[a_i,b_i]$,
\item[(b)]
$f_i(a_i)\in\ZZ$ and $f_i(b_i)-f_i(a_i)=k_i$ for some integer $k_i\in\ZZ$,
\item[(c)]
$f_i\in\C^2([a_i,b_i])$ with derivative $f_i'(q)\neq 0$ for
all $q\in[a_i,b_i]$.
\end{itemize}
Let us define the constant
\begin{equation}
\gamma_i = \sign(k_i) \max_{q\in[a_i,b_i]} |f_i'(q)| .
\end{equation}

\begin{thm}
Suppose either of the following conditions is satisfied,
\begin{itemize}
\item[(i)]
$\gamma_1\gamma_2<0$;
\item[(ii)]
$|k_1|,|k_2|\geq 2$ and $\gamma_1 \gamma_2 > C_0
\approx 17.24445$.
\end{itemize}
Then the map \eqref{ltm} acts ergodically
(with respect to Lebesgue measure $\mu$) on the domain
\begin{equation}
\scrD_0=\big\{ (p,q)\in\TT^2 : p\in[a_1,b_1] \big\}
\cup
\big\{ (p,q)\in\TT^2 : q\in[a_2,b_2] \big\} .
\end{equation}
\end{thm}

The proofs of the two parts (i) and (ii) of this statement
are due to Burton and Easton \cite{Burton80}, and Przytycki \cite{Przytycki83},
respectively. Both \cite{Burton80} and \cite{Przytycki83}
in fact establish the Bernoulli property for the action of $\Phi$ on $\scrD_0$
under conditions (i), (ii). We expect that these properties hold
under weaker conditions, e.g., for smaller values of $C_0$.
The continuity of the map at the lines $p=a_1,b_1$ and $q=a_2,b_2$,
assumed in condition (b), is probably also not necessary.

For the linked twist map $\Phi$ used in Figures \ref{figQE}--\ref{figWeyl}
we have chosen
\begin{equation}
[a_1,b_1]=[a_2,b_2]=[\tfrac12,1],
\end{equation}
\begin{equation}
f_1(p)=
\begin{cases}
-2p & \text{if } p\in [\tfrac12,1] \\
0   & \text{if } p\notin [\tfrac12,1] ,
\end{cases}
\qquad
V_1(p)=
\begin{cases}
p^2 & \text{if } p\in [\tfrac12,1] \\
0   & \text{if } p\notin [\tfrac12,1] ,
\end{cases}
\end{equation}
and
\begin{equation}
f_2(q)=
\begin{cases}
2q & \text{if } q\in [\tfrac12,1] \\
0   & \text{if } q\notin [\tfrac12,1] ,
\end{cases}
\qquad
V_2(q)=
\begin{cases}
-q^2 & \text{if } q\in [\tfrac12,1] \\
0   & \text{if } q\notin [\tfrac12,1] .
\end{cases}
\end{equation}
More explicitly, this particular map
$\Phi$ is obtained by first applying the twist
\begin{equation}
\begin{pmatrix} p\\ q \end{pmatrix}
\mapsto
\begin{cases}
\begin{pmatrix} 1 & 2  \\ 0 & 1 \end{pmatrix}
\begin{pmatrix} p \\ q \end{pmatrix} & \text{if } q\in [\tfrac12,1] \\[15pt]
\begin{pmatrix} 1 & 0  \\ 0 & 1 \end{pmatrix}
\begin{pmatrix} p\\ q \end{pmatrix} &  \text{if } q\not\in [\tfrac12,1],
\end{cases}
\end{equation}
followed by
\begin{equation}
\begin{pmatrix} p\\ q \end{pmatrix}
\mapsto
\begin{cases}
\begin{pmatrix} 1 & 0  \\ 2 & 1 \end{pmatrix}
\begin{pmatrix} p \\ q \end{pmatrix} & \text{if } p\in [\tfrac12,1] \\[15pt]
\begin{pmatrix} 1 & 0  \\ 0 & 1 \end{pmatrix}
\begin{pmatrix} p\\ q \end{pmatrix} &  \text{if } p\not\in [\tfrac12,1] .
\end{cases}
\end{equation}
Clearly, this $\Phi$ satisfies the above conditions (a)--(c) and (i).

\subsection{Discretised linked twist maps}\label{appDis}

This section illustrates that the results of this paper may be
applied to problems outside quantum mechanics, such as
the discretization of maps on the torus, cf.~\cite{Bosio00,Benatti03}
and references therein.
A {\em discretization} of $\Phi:\TT^2\to\TT^2$ is defined as the
invertible\footnote{Invertability
is not necessarily required in general discretization schemes.
We assume it here to obtain a unitary representation.} map
$\Phi_M: \TT^2_M \to \TT^2_M$ with $\TT^2_M:=(M^{-1}\ZZ/\ZZ)^2$,
where $\Phi_M$ is chosen in such a way that
\begin{equation}
\lim_{M\to\infty} \sup_{\xi\in\TT^2_M} d\big(\Phi_M(\xi),\Phi(\xi)\big) =0 ;
\end{equation}
$d(\,\cdot\,,\,\cdot\,)$ denotes the Riemannian distance on $\TT^2$.
The discretised map induces a permutation matrix $U_N(\Phi):\CC^N\to\CC^N$
with $N=M^2$ defined by
\begin{equation}
[U_N(\Phi)\psi](\xi)
= \psi\circ\Phi_M^{-1}(\xi) , \qquad \xi\in \TT_M^2
\end{equation}
(we represent vectors $\psi\in\CC^N$ as functions
$\psi:\TT^2_M \to \CC$).

The ``quantum'' observables required in Axiom \ref{quantass}
are simply defined as multiplication operators,
\begin{equation}
[\Op_N(a)\psi](\xi)=a(\xi)\, \psi(\xi)
\end{equation}
and therefore trivially satisfy Axioms \ref{quantass} (a) and
(b). As to (c),
\begin{equation}
\frac1N \Tr\Op_N(a) = \frac{1}{M^2} \sum_{m_1,m_2=0}^{M-1}
a\left(\frac{m_1}{M},\frac{m_2}{M}\right)
=
\int_{\TT^2} a \, d\mu +O(M^{-1}).
\end{equation}
It is easily checked that
\begin{equation}
U_N(\Phi)^{-1} \Op_N(a) U_N(\Phi) = \Op_N(a \circ \Phi_M)
\end{equation}
and thus
\begin{align}
\|U_N(\Phi)^{-1} \Op_N(a) U_N(\Phi) - \Op_N(a \circ \Phi)\|
&=\|\Op_N(a \circ \Phi_M)- \Op_N(a \circ \Phi)\|   \\
&\leq \sup_{\xi\in\TT^2_M} |a(\Phi_M(\xi))-a(\Phi(\xi))| \\
&\ll\sup_{\xi\in\TT^2_M} d\big(\Phi_M(\xi),\Phi(\xi)\big) ,
\end{align}
and hence Axiom \ref{corresp} is satisfied. Condition \eqref{spezi} follows
from a similar argument, with $\alpha_\scrD=0$.

A concrete discretization of a linked twist map is obtained
by replacing each twist map $\Psi_f$ by $\Psi_{f_M}$,
with $f_M(q)=M^{-1}[M f(q)]$, where $[x]$ denotes the integer
part of $x$. Note that $\R$ preserves $\TT^2_M$ and therefore requires no
further discretization.
In this case
\begin{equation}
\sup_{\xi\in\TT^2_M} d\big(\Phi_M(\xi),\Phi(\xi)\big) = O(M^{-1}) .
\end{equation}

There is a simple geometric interpretation of the spectrum
of $U_N(\Phi)$. The map $\Phi_M$ represents a permutation
of $N=M^2$ elements, which can be written as a product of,
say, $\nu$ cycles $C_i$ of length $\ell_i$, $i=1,\ldots,\nu$.
Each cycle corresponds to a periodic orbit of period $\ell_i$
for the action of $\Phi_M$ on $\TT^2_M$. Let $\xi_i$ be an arbitrary
point on $C_i$. An orthonormal basis of
eigenvectors of $U_N(\Phi)$ is then given by
the functions
\begin{equation}
\varphi_{ij}(\xi)=
\begin{cases}
\ell_i^{-1/2}
e( jk /\ell_i)& \text{if $\xi=\Phi_M^{-k}(\xi_i)$ for some $k=0,\ldots,\ell_i-1$,}
\\
0 & \text{otherwise,}
\end{cases}
\end{equation}
with eigenvalue $\lambda_{ij}=e( j /\ell_i)$,
where $i$ runs over the cycles and $j$
over the integers $0,1,\ldots,\ell_i-1$.
Hence in particular $|\varphi_{ij}(\xi)|^2=1/\ell_i$
for every $\xi$ on the periodic
orbit, and $|\varphi_{ij}(\xi)|^2=0$ otherwise.

Due to the commutativity of the observables $\vecOp(a)$,
the proofs of some of the statements
in later sections may be simplified---especially those in Section \ref{secLoc}.

\section{Mollified characteristic functions}

Let us now return to the general framework of Section \ref{secSetup}.
Consider the characteristic function $\chi_\scrD$ of a domain
$\scrD\subset\scrM$ with boundary of Minkowski content zero.
An {\em $\epsilon$-mollified characteristic function}
$\schi_\scrD\in\C^\infty(\scrM)$ has values in $[0,1]$ and
$\schi_\scrD(x)=\chi_\scrD(x)$ on a set of $x$ of measure $1-\epsilon$.
Since $\scrD$ has boundary of Minkowski content zero,
we can construct such a smoothed
function for any $\epsilon>0$. Furthermore we are able to construct
$\epsilon$-mollified $\schi_\scrD$ whose support is either contained
in $\scrD$, or whose support contains $\scrD$, again for any $\epsilon>0$.
Note that if $\schi_\scrD$ is $\epsilon$-mollified, so is $\schi_\scrD^n$
for any $n\in\NN$ with the same $\epsilon$.

After mollification, we may associate with a characteristic function
$\chi_\scrD$ a quantum observable $\Op_N(\schi_\scrD)$. Since
$\Op_N(\schi_\scrD)$ is in general not hermitian, it is sometimes more
convenient to consider the symmetrised version, the positive definite
hermitian matrix
\begin{equation}\label{pd}
\Op_N^{\sym}(\schi_\scrD):=
\Op_N(\schi_\scrD^{1/2})\Op_N(\schi_\scrD^{1/2})^\dagger.
\end{equation}
Note that $\schi_\scrD^{1/2}\in\C^\infty(\scrM)$ since $\schi_\scrD\geq 0$.
Furthermore, we have
\begin{equation}
\vecOp^{\sym}(\schi_\scrD) \sim
\vecOp(\schi_\scrD) .
\end{equation}

The following proposition describes the distribution of eigenvalues
of $\Op_N^{\sym}(\schi_\scrD)$, and suggests that
the operator may be viewed as an approximate projection operator
onto a subspace of dimension $\sim N\times \mu(\scrD)$.

Consider a sequence $\vecJ:=\{J_N\}_{N\in\scrI}$ of sets
$J_N\subset \{1,\ldots,N\}$.
The quantity
\begin{equation}
\Delta(\vecJ):=\lim_{N\to\infty} \frac{\# J_N}{N} ,
\end{equation}
provided the limit exists, is called the {\em density of} $\vecJ$.

\begin{prop}\label{proppy}
Suppose $\schi_\scrD$ is an $\epsilon$-mollified characteristic function, and
suppose $\mu_j\geq 0$ $(j=1,\ldots,N)$ are the eigenvalues of
$\Op^{\sym}_N(\schi_\scrD)$.
Then there are set sequences $\vecJ:=\{J_N\}_{N\in\scrI}$
and $\vecJ':=\{J_N'\}_{N\in\scrI}$ with densities
\begin{equation}
\Delta(\vecJ)=\mu(\scrD)+O(\epsilon^{1/3}), \qquad
\Delta(\vecJ')=1-\mu(\scrD)+O(\epsilon^{1/3}),
\end{equation}
such that
\begin{itemize}
\item[(i)]
$\mu_{j}= 1+O(\epsilon^{1/3})$ for all $j\in J_N$;
\item[(ii)]
$\mu_{j}= O(\epsilon^{1/3})$ for all $j\in J_N'$.
\end{itemize}
\end{prop}

\begin{proof}
By Axiom \ref{quantass}, we have for every fixed integer $n\geq 1$,
\begin{align}
\frac1N \Tr \left[\Op^{\sym}_N(\schi_\scrD)^n \right]
&=\frac1N \Tr \Op^{\sym}_N(\schi_\scrD^{n}) + o_{\epsilon,n}(1) \\
&=\int_\scrM \schi_\scrD^{n} d\mu + o_{\epsilon,n}(1) \\
&=  \mu(\scrD) +O(\epsilon)+o_{\epsilon,n}(1) ,
\end{align}
where $O(\epsilon)$ does not depend on $N$ and $n$.
This implies for every $n\geq 1$,
\begin{equation}\label{moment}
\lim_{N\to\infty} \frac1N
\Tr \left[\Op^{\sym}_N(\schi_\scrD)^n \right]
=
\lim_{N\to\infty} \frac1N \sum_{j=1}^N \mu_j^n
=
\mu(\scrD) +O(\epsilon) .
\end{equation}
Therefore
\begin{equation}
\lim_{N\to\infty} \frac1N \sum_{j=1}^N (\mu_j^2-\mu_j)^{2} = O(\epsilon)
\end{equation}
and thus
\begin{equation}\label{Hj}
\lim_{N\to\infty} \frac1N \sum_{j\in H_N}
(\mu_j-1)^{2} = O(\epsilon) .
\end{equation}
\begin{equation}\label{Hj0}
\lim_{N\to\infty} \frac1N \sum_{j \notin H_N}
\mu_j^{2} = O(\epsilon) ,
\end{equation}
where $H_N=\{ j : \mu_j\geq 1/2 \}$. By Chebyshev's inequality,
\eqref{Hj0} implies that
\begin{equation}\label{Cheby}
\lim_{N\to\infty} \frac1N \#\{ j \notin H_N :
\mu_j^2 > \gamma \} = O(\epsilon/\gamma) ,
\end{equation}
for any $\gamma>0$. This yields the bound
\begin{equation}\label{Hj0first}
\lim_{N\to\infty} \frac1N \sum_{j \notin H_N}
\mu_j = O\big(\gamma^{1/2}+\epsilon/\gamma\big),
\end{equation}
since $0\leq \mu_j< 1/2$.
So
\begin{align}
\lim_{N\to\infty} \frac1N  \sum_{j\in H_N} 1
& = \lim_{N\to\infty} \frac1N \sum_{j\in H_N} \left[ (\mu_j-1)^{2} -
\mu_j^2 + 2 \mu_j \right] \\
& = \lim_{N\to\infty} \frac1N \sum_{j\in H_N} \left[-\mu_j^2 + 2\mu_j \right]
+ O(\epsilon) \\
& = \lim_{N\to\infty} \frac1N \sum_{j=1}^N \left[-\mu_j^2 + 2\mu_j \right]
+ O(\epsilon)+ O\big(\gamma^{1/2}+\epsilon/\gamma\big) \\
& = \mu(\scrD) + O(\epsilon^{1/3}) ,
\end{align}
if we choose $\gamma=\epsilon^{2/3}$,
and hence the corresponding set sequence
$\vecH:=\{H_N\}_{N\in\scrI}$
has density $\Delta(\vecH)=\mu(\scrD) +O(\epsilon^{1/3})$.
Once more in view of Chebyshev's inequality, \eqref{Hj} implies
\begin{equation}
\lim_{N\to\infty} \frac1N \#\{ j \in H_N :
(\mu_j-1)^{2} > \delta \} = O(\epsilon/\delta).
\end{equation}
Choosing $\delta=\epsilon^{2/3}$,
this means that for a subsequence of $j\in H_N$
of density $\mu(\scrD)+O(\epsilon^{1/3})$
we have $\mu_j=1+O(\epsilon^{1/3})$. The corresponding result for
$j\notin H_N$ follows by the same argument from \eqref{Cheby}.
\end{proof}

\section{Trace asymptotics and Weyl's law}\label{secTra}

The following proposition is the key tool to understand the
distribution of eigenvalues of $U_N(\Phi)$.

\begin{prop}[Trace asymptotics] \label{traceasymp}
Suppose $\Phi:\scrM\to\scrM$ is piecewise smooth, and
\begin{itemize}
\item[(a)]
$\Phi^n\big|_{\scrD'}=\id$ on some set $\scrD'\subset\scrM$
with boundary of Minkowski content zero;
\item[(b)]
the fixed points of $\Phi^n$ on $\scrD:=\scrM-\scrD'$ form a set
of Minkowski content zero;
\item[(c)]
there is a constant $\alpha_{\scrD'}\in\RR$ such that for any
$a\in\C^\infty(\scrM)$ with compact support
contained in $\scrD'$ and
the domains of continuity of $\Phi^n$, we have
\begin{equation*}
\vecU(\Phi)^{n} \vecOp(a) \sim e(n\alpha_{\scrD'})\, \vecOp(a) .
\end{equation*}
\end{itemize}
Then
\begin{equation}
\lim_{N\to\infty} \frac1N \Tr U_N(\Phi)^n = e(n\alpha_{\scrD'})\, \mu(\scrD').
\end{equation}
\end{prop}

\begin{proof}
Given any $\epsilon>0$, we can find an integer $R$ and a
partition of unity on $\scrM$
by $\epsilon$-mollified characteristic functions,
\begin{equation}
1= \schi_{\text{bad}}(\xi)+ \schi_{\scrD'}(\xi) + \sum_{r=1}^R \schi_r(\xi)
\quad \forall \xi\in\scrM
\end{equation}
with the properties
\begin{itemize}
\item[(i)]
the interior of the support of $\schi_{\text{bad}}$ contains
the domains of discontinuity of $\Phi^\nu$ in $\scrM$ for all $\nu=1,\ldots,n$,
and all fixed points of $\Phi^n$ in
$\scrD$, and is chosen small enough so that
$\int \schi_{\text{bad}} d\mu < \epsilon$;
\item[(ii)]
the support of $\schi_{\scrD'}$ is contained
in $\scrD'$ and the domains of continuity
of $\Phi^\nu$ for all $\nu=1,\ldots,n$, so that
$\mu(\scrD')-\int \schi_{\scrD'} d\mu < \epsilon$
and all fixed points of $\Phi^n$ are contained in the interior
of the set $\supp\schi_{\text{bad}} \cup \supp \schi_{\scrD'}$;
\item[(iii)]
the support of $\schi_r$, with $r=1,\ldots,R$, is chosen small enough, so that
$\supp\schi_r\cap\Phi^n(\supp\schi_r)=\emptyset$ for all $\xi\in\scrM$.
\end{itemize}
Properties (i) and (ii) are possible since the fixed points in $\scrD$
and the discontinuities
form sets of Minkowski content zero. To achieve (iii) note that
the closure of
$\scrK=\scrM-(\supp\schi_{\text{bad}} \cup \supp \schi_{\scrD'})$
does not contain any fixed points, and $\Phi$ is continuous on $\scrK$.
Hence there is a sufficiently small radius $\eta=\eta(\epsilon)$ such that
for all balls $\scrB_\eta\subset\scrK$
we have $\scrB_\eta\cap\Phi^n(\scrB_\eta)=\emptyset$.

By the linearity of $\Op$, we have
\begin{equation}\label{triple}
\Tr U_N(\Phi)^n
=
\Tr [U_N(\Phi)^n \Op_N(\schi_{\text{bad}})]
+ \Tr [U_N(\Phi)^n \Op_N(\schi_{\scrD'})] +
\sum_{r=1}^R \Tr [U_N(\Phi)^n \Op_N(\schi_r)] .
\end{equation}

We begin with the first term on the right hand side:
\begin{equation}
\Tr [U_N(\Phi)^n \Op_N(\schi_{\text{bad}})]
=\Tr [U_N(\Phi)^n
\Op^{\sym}_N(\schi_{\text{bad}})]
+o_\epsilon(N),
\end{equation}
with the symmetrised $\Op^{\sym}_N(\schi_{\text{bad}})$
as defined in \eqref{pd}.
Suppose $\psi_j$ and $\mu_j\geq 0$
are the (normalised) eigenstates and
eigenvalues of $\Op^{\sym}_N(\schi_{\text{bad}})$.
Then
\begin{multline}
|\Tr [U_N(\Phi)^n \Op^{\sym}_N(\schi_{\text{bad}})]|
= \bigg| \sum_{j=1}^N \mu_j \langle
\psi_j, U_N(\Phi)^n \psi_j \rangle \bigg|
\leq \sum_{j=1}^N \mu_j \\
=
\Tr\Op^{\sym}_N(\schi_{\text{bad}})
= \Tr\Op_N(\schi_{\text{bad}})+o_\epsilon(N)
=N O(\epsilon)+o_\epsilon(N).
\end{multline}

Since by condition (c),
$\vecU(\Phi)^n \vecOp(\schi_{\scrD'}) \sim e(n\alpha_{\scrD'})
\vecOp(\schi_{\scrD'})$,
we find for the second term on the right hand side of \eqref{triple}
\begin{align}
\Tr [U_N(\Phi)^n \Op_N(\schi_{\scrD'})]
& =
e(n\alpha_{\scrD'}) \Tr \Op_N(\schi_{\scrD'}) +o_\epsilon(N) \\
& =
N e(n\alpha_{\scrD'}) \int \schi_{\scrD'} d\mu +o_\epsilon(N)\\
& =N e(n\alpha_{\scrD'}) \left\{ \mu(\scrD')+O(\epsilon)\right\}
+o_\epsilon(N) ,
\end{align}
where we have used (ii) in the last step.

For the last term in the sum \eqref{triple}
we have
\begin{equation}
\vecU(\Phi)^n \vecOp(\schi_r)
\sim \vecU(\Phi)^n \vecOp(\schi_r^{1/2}) \vecOp(\schi_r^{1/2})
\sim \vecOp(\schi_r^{1/2}\circ\Phi^{-n}) \vecU(\Phi)^n \vecOp(\schi_r^{1/2})
\end{equation}
so
\begin{align}
\Tr [U_N(\Phi)^n \Op_N(\schi_r)]
& = \Tr[\Op_N(\schi_r^{1/2}\circ\Phi^{-n}) U_N(\Phi)^n \Op_N(\schi_r^{1/2}) ]
+o_\epsilon(N) \\
& = \Tr[\Op_N(\schi_r^{1/2}) \Op_N(\schi_r^{1/2}\circ\Phi^{-n}) U_N(\Phi)^n  ]
+o_\epsilon(N) \\
& = \Tr[\Op_N(\schi_r^{1/2}\cdot\schi_r^{1/2}\circ\Phi^{-n}) U_N(\Phi)^n  ]
+o_\epsilon(N) \\
& = o_\epsilon(N)
\end{align}
since $\schi_r^{1/2}\cdot\schi_r^{1/2}\circ\Phi^{-n}=0$ in view of (iii).
Therefore
\begin{equation}
\Tr U_N(\Phi)^n=N e(n\alpha_{\scrD'}) \left\{ \mu(\scrD')+O(\epsilon)\right\}
+o_\epsilon(N),
\end{equation}
i.e.,
\begin{equation}
\lim_{N\to\infty} \frac1N \Tr U_N(\Phi)^n = e(n\alpha_{\scrD'})\, \mu(\scrD')
+O(\epsilon),
\end{equation}
which holds for every arbitrarily small $\epsilon>0$. This concludes the proof.
\end{proof}

\begin{thm}[Weyl's law]\label{weyls}
Suppose $\Phi:\scrM\to\scrM$ is piecewise smooth, and
\begin{itemize}
\item[(a)]
there are disjoint sets $\scrD_\nu$ $(\nu=1,2,3,\ldots)$
with boundary of Minkowski content zero, on which
$\Phi$ is periodic with period $n_\nu\geq 1$, i.e.,
\begin{equation*}
\Phi^{n_\nu}\big|_{\scrD_\nu}=\id ,
\end{equation*}
and, for every $0<|n|<n_\nu$, the fixed points of $\Phi^n$
on $\scrD_\nu$ form a set of Minkowski content zero;
\item[(b)]
for every $n\neq 0$, the fixed points of $\Phi^n$ on
$\scrD_0=\scrM-\bigcup_{\nu=1}^\infty \scrD_\nu$ form a set
of Minkowski content zero;
\item[(c)]
there is a constant $\alpha_{\scrD_\nu}\in\RR$
such that for any $a\in\C^\infty(\scrM)$ with compact support
contained in $\scrD_\nu$ and
the domains of continuity of $\Phi,\Phi^2,\ldots,\Phi^{n_\nu}$, we have
\begin{equation*}
\vecU(\Phi)^{n_\nu} \vecOp(a) \sim e(n_\nu \alpha_{\scrD_\nu})\, \vecOp(a) .
\end{equation*}
\end{itemize}
Then, for every continuous function $h:\SS^1=\RR/\ZZ\to\CC$,
\begin{equation}
\lim_{N\to\infty} \frac1N \sum_{j=1}^N h(\theta_j) =
\sum_{\nu=0}^\infty \mu(\scrD_\nu)
\int_{\SS^1} h(\theta)\, \rho_\nu(\theta)\, d\theta
\end{equation}
where
\begin{equation}
\rho_0(\theta) = 1 ,
\end{equation}
and, for $\nu\geq 1$,
\begin{equation}
\rho_\nu(\theta) =
\frac{1}{n_\nu} \sum_{k=0}^{n_\nu-1}
\delta_{\SS^1}\bigg(\theta-\frac{k}{n_\nu}-\alpha_{\scrD_\nu}\bigg) .
\end{equation}
\end{thm}
Here
\begin{equation}
\delta_{\SS^1}(\theta) = \sum_{m\in\ZZ} \delta(\theta+m)
\end{equation}
denotes the periodicised Dirac distribution.

\begin{proof}
For every $\nu$ such that $n_\nu$ divides $n$ we have
$\Phi^{n}\big|_{\scrD_\nu}=\id$. A simple modification of the proof of
Proposition \ref{traceasymp} yields therefore, for $n\neq 0$,
\begin{equation}\label{turn}
\lim_{N\to\infty} \frac1N \Tr U_N(\Phi)^n =
\sum_{\substack{\nu=1\\ n_\nu|n}}^\infty
e(n\alpha_{\scrD_\nu}) \, \mu(\scrD_\nu) .
\end{equation}
The only difference in the proof is that $\scrD'$ is divided into
the domains $\scrD_1,\scrD_2,\ldots$ with different integration constants
$\alpha_{\scrD_\nu}$. For every $\epsilon>0$ there is a
$K=K_\epsilon$ such that
\begin{equation}\label{sixfour0}
\mu\bigg(\bigcup_{\nu=K+1}^\infty \scrD_\nu\bigg) < \epsilon .
\end{equation}
Hence one effectively deals with only finitely many domains
$\scrD_1,\ldots,\scrD_K$ and shows,
following the steps in the proof of Proposition \ref{traceasymp}, that
\begin{equation}
\lim_{N\to\infty} \frac1N \Tr U_N(\Phi)^n =
\sum_{\substack{\nu=1\\ n_\nu|n}}^K
e(n\alpha_{\scrD_\nu}) \, \mu(\scrD_\nu) + O(\epsilon),
\end{equation}
which in turn yields \eqref{turn}.

Let us first assume that the test function $h$ has
only finitely many non-zero Fourier coefficients, i.e.,
\begin{equation}\label{finiteFS}
h(\theta)=\sum_{n\in\ZZ} \widehat h(n) e(n\theta)
\end{equation}
is a finite sum.
We then have
\begin{align}
\lim_{N\to\infty} \frac1N \sum_{j=1}^N h(\theta_j)
& =
\lim_{N\to\infty} \frac1N \sum_{n\in\ZZ} \widehat h(n) \Tr U_N(\Phi)^n \\ & =
\widehat h(0)+
\sum_{n\in\ZZ-\{0\}} \widehat h(n) \sum_{\substack{\nu=1\\ n_\nu|n}}^\infty
e(n\alpha_{\scrD_\nu})\, \mu(\scrD_\nu) \\ & =
\widehat h(0)+
\sum_{\nu=1}^\infty  \mu(\scrD_\nu)
\sum_{\substack{n\in\ZZ-\{0\} \\ n_\nu|n}}
\widehat h(n) e(n\alpha_{\scrD_\nu}).
\end{align}
Since
\begin{equation}
\sum_{\substack{n\in\ZZ \\ n_\nu|n}} \widehat h(n) e(n\alpha_{\scrD_\nu})
=
\frac{1}{n_\nu} \sum_{k=0}^{n_\nu-1}
\sum_{n\in\ZZ} \widehat h(n) e(n\alpha_{\scrD_\nu})\,
e\bigg(\frac{k n}{n_\nu}\bigg)
=
\frac{1}{n_\nu} \sum_{k=0}^{n_\nu-1}
h\bigg(\frac{k}{n_\nu}+\alpha_{\scrD_\nu}\bigg)
\end{equation}
we obtain
\begin{equation}\label{so}
\lim_{N\to\infty} \frac1N \sum_{j=1}^N h(\theta_j)
=\left\{ \mu(\scrD_0)\,  \widehat h(0)+
\sum_{\nu=1}^\infty
\frac{\mu(\scrD_\nu)}{n_\nu}
\sum_{k=0}^{n_\nu-1} h\bigg(\frac{k}{n_\nu}+\alpha_{\scrD_\nu}\bigg) \right\} ,
\end{equation}
which proves the theorem for $h$ with finite Fourier series.
We now extend this result to test functions $h\in\C^1(\SS^1)$.
Let
\begin{equation}
h_K(\theta)=\sum_{\substack{n\in\ZZ\\ |n|\leq K}} \widehat h(n) e(n\theta)
\end{equation}
be the truncated Fourier series. Since $h\in\C^1(\SS^1)$, its
Fourier series converges absolutely and uniformly
and hence, for any $\epsilon>0$,
there is a $K$ such that
$h_K(\theta) - \epsilon \leq h(\theta) \leq h_K(\theta) + \epsilon$
for all $\theta\in\SS^1$.
By \eqref{so}, the limits of the left and right hand side of
\begin{equation}
\frac1N \sum_{j=1}^N h_K(\theta_j) -\epsilon
\leq
\frac1N \sum_{j=1}^N h(\theta_j)
\leq
\frac1N \sum_{j=1}^N h_K(\theta_j) +\epsilon
\end{equation}
exist and differ by less than $2\epsilon$, hence \eqref{so}
holds also for the current $h$. The extension of \eqref{so}
to $h$ in $\C(\SS^1)$ is achieved by the same argument, i.e.,
by approximating $h$ pointwise by functions $h_\epsilon\in\C^1(\SS^1)$
so that
$h_\epsilon(\theta) - \epsilon \leq h(\theta)
\leq h_\epsilon(\theta) + \epsilon$.
\end{proof}

\section{Generalised Weyl's law}\label{secGen}

\begin{prop}[Generalised trace asymptotics] \label{traceasymp2}
Choose $\Phi$ and $U_N(\Phi)$ as in Proposition \ref{traceasymp}.
Then for every $a\in\C^\infty(\scrM)$ and $n\neq 0$,
\begin{equation}\label{GTA}
\lim_{N\to\infty} \frac1N \Tr[\Op_N(a)U_N(\Phi)^n] =
e(n\alpha_{\scrD'}) \int_{\scrD'} a\,d\mu .
\end{equation}
\end{prop}

\begin{proof}
By linearity of the relation \eqref{GTA} we may
assume without loss of generality that $a$ is real and $\min_\xi a(\xi)\geq 0$.
This implies that $a^{1/2}\in\C^\infty(\scrM)$.
Analogously to the proof of Proposition \ref{traceasymp}, we have
\begin{multline}\label{triple2}
\Tr[\Op_N(a)U_N(\Phi)^n]
=
\Tr [U_N(\Phi)^n \Op_N(\schi_{\text{bad}}\cdot a)] \\
+ \Tr [U_N(\Phi)^n \Op_N(\schi_{\scrD'}\cdot a)]
+ \sum_{r=1}^R \Tr [U_N(\Phi)^n \Op_N(\schi_r\cdot a)] + o_{\epsilon}(N).
\end{multline}
The proof is concluded in the same way as the
proof of Proposition \ref{traceasymp}, with all mollified characteristic
functions $\schi$ replaced by $\schi\cdot a$.
\end{proof}

\begin{thm}[Generalised Weyl's law]\label{weyls2}
Choose $\Phi$ and $U_N(\Phi)$ as in Theorem \ref{weyls}.
Let $\varphi_j\in\CC^N$ $(j=1,\ldots,N)$ be
an orthonormal basis of eigenstates of $U_N(\Phi)$,
with corresponding eigenphases $\theta_j\in\SS^1$.
Then, for every $a\in\C^\infty(\scrM)$ and
every continuous function $h:\SS^1\to\CC$,
\begin{equation}\label{weyls2eq}
\lim_{N\to\infty} \frac1N \sum_{j=1}^N h(\theta_j)
\langle \Op_N(a) \varphi_j,\varphi_j \rangle =
\sum_{\nu=0}^\infty \int_{\scrD_\nu} a\, d\mu \;
\int_{\SS^1} h(\theta)\, \rho_\nu(\theta)\, d\theta .
\end{equation}
\end{thm}

\begin{proof}
We may assume again without loss of generality that $a$ is real and
$\min_\xi a(\xi)\geq 0$.
In view of Proposition \ref{traceasymp2} and the proof of Theorem \ref{weyls}
we have for every $h_K$ with finite Fourier expansion (as in \eqref{finiteFS})
\begin{equation}\label{hK}
\lim_{N\to\infty} \frac1N \sum_{j=1}^N h_K(\theta_j)
\langle \Op_N(a) \varphi_j,\varphi_j \rangle =
\sum_{\nu=0}^\infty \int_{\scrD_\nu} a\, d\mu \;
\int_0^1 h_K(\theta)\, \rho_\nu(\theta)\, d\theta .
\end{equation}
For any $h\geq 0$ we have
\begin{multline}\label{poseq}
\left| \sum_{j=1}^N h(\theta_j)
\langle \Op_N(a) \varphi_j,\varphi_j \rangle
-
\sum_{j=1}^N h(\theta_j)
\| \Op_N(a^{1/2}) \varphi_j\|^2  \right| \\
 \leq
\sup h \left| \Tr[\Op_N(a)-\Op_N(a^{1/2})\Op_N(a^{1/2})^\dagger] \right|
= o(N) \sup h .
\end{multline}
Hence \eqref{hK} is equivalent to
\begin{equation}\label{hK2}
\lim_{N\to\infty} \frac1N \sum_{j=1}^N h_K(\theta_j)
\| \Op_N(a^{1/2}) \varphi_j \|^2 =
\sum_{\nu=0}^\infty \int_{\scrD_\nu} a\, d\mu \;
\int_0^1 h_K(\theta)\, \rho_\nu(\theta)\, d\theta .
\end{equation}
We now use the same approximation argument as in the proof of
Theorem \ref{weyls},
for $h\in\C^1(\SS^1)$. Given any $\epsilon$,
there is a $K$ such that
$h_K(\theta) - \epsilon \leq h(\theta) \leq h_K(\theta) + \epsilon$
for all $\theta\in\SS^1$.
The limits of the left and right hand side of
\begin{align}
\frac1N \sum_{j=1}^N [h_K(\theta_j)-\epsilon] \| \Op_N(a^{1/2}) \varphi_j \|^2
& \leq
\frac1N \sum_{j=1}^N h(\theta_j) \| \Op_N(a^{1/2}) \varphi_j \|^2 \\
& \leq
\frac1N \sum_{j=1}^N [h_K(\theta_j) +\epsilon]\| \Op_N(a^{1/2}) \varphi_j \|^2
\end{align}
differ by less than
\begin{align}
2\epsilon \sup h_K \lim_{N\to\infty}
\frac1N \sum_{j=1}^N\| \Op_N(a^{1/2}) \varphi_j \|^2
& \leq
2\epsilon \sup h_K \lim_{N\to\infty} \frac1N
\Tr[\Op_N(a^{1/2})\Op_N(a^{1/2})^\dagger]\\
& =
2\epsilon \sup h_K \int_\scrM a\, d\mu
\end{align}
which can be arbitrarily small for $\epsilon\to 0$. Thus
\begin{equation}\label{h2}
\lim_{N\to\infty} \frac1N \sum_{j=1}^N h(\theta_j)
\| \Op_N(a^{1/2}) \varphi_j \|^2 =
\sum_{\nu=0}^\infty \int_{\scrD_\nu} a\, d\mu \;
\int_0^1 h(\theta)\, \rho_\nu(\theta)\, d\theta .
\end{equation}
A similar approximation argument shows that \eqref{h2} holds also
for all continuous $h$. In view of \eqref{poseq}, the relation \eqref{h2}
is equivalent to \eqref{weyls2eq}. The assumption $h\geq 0$ can be
removed by using the linearity of \eqref{weyls2eq} in $h$.
\end{proof}

\section{Localization}\label{secLoc}

The sequence $\vecpsi:=\{ \psi_N \}_{N\in\scrI}$ of vectors
$\psi_N\in\CC^N-\{\vecnull\}$ is said to be
{\em semiclassically localised in the domain} $\scrD$ if
for every $a\in\C^\infty(\scrM)$ with $a|_\scrD=0$,
we have
\begin{equation}
\lim_{N\to\infty} \frac{\langle \Op_N(a)
\psi, \psi \rangle}{\|\psi\|^2} =0.
\end{equation}

\begin{thm} \label{locthm}
Choose $\Phi$ and $U_N(\Phi)$ as in Theorem \ref{weyls}, and let
$\varphi_1,\ldots,\varphi_N\in\CC^N$ be
an orthonormal basis of eigenstates of $U_N(\phi)$.
Then there are set sequences $\vecJ:=\{J_N\}_{N\in\scrI}$
and $\vecJ':=\{J_N'\}_{N\in\scrI}$ with densities
\begin{equation}
\Delta(\vecJ)=\mu(\scrD_0), \qquad
\Delta(\vecJ')=1-\mu(\scrD_0),
\end{equation}
such that
\begin{itemize}
\item[(i)]
for any $a\in\C^\infty(\scrM)$  with $a|_{\scrD_0}=0$,
\begin{equation*}
\lim_{N\to\infty} \frac1N \sum_{j\in J_N}
\left| \langle \Op_N(a) \varphi_j , \varphi_j \rangle \right|^2 = 0 ;
\end{equation*}
\item[(ii)]
for any $a\in\C^\infty(\scrM)$  with $a|_{\scrM-\scrD_0}=0$,
\begin{equation*}
\lim_{N\to\infty} \frac1N \sum_{j\in J_N'}
\left| \langle \Op_N(a) \varphi_j , \varphi_j \rangle \right|^2 = 0 .
\end{equation*}
\end{itemize}
\end{thm}

\begin{proof}
For any given $\epsilon>0$, there is a constant $K=K_\epsilon$ such that
\begin{equation}\label{sixfour}
\mu\bigg(\bigcup_{\nu=K+1}^\infty \scrD_\nu\bigg) < \epsilon .
\end{equation}
Consider the following subset of $\SS^1=\RR/\ZZ$,
\begin{equation}
\Theta_\epsilon
:=\bigcup_{\nu=1}^K \bigcup_{k=0}^{n_\nu-1}
\left[\frac{k}{n_\nu}+\alpha_{\scrD_\nu}-\frac{\epsilon}{K n_\nu},
\frac{k}{n_\nu}+\alpha_{\scrD_\nu}+\frac{\epsilon}{Kn_\nu} \right]+ \ZZ.
\end{equation}
We may construct a continuous function
$h=h_\epsilon$ with values in $[0,1]$ such that
\begin{equation}
h(\theta)=
\begin{cases}
1 & \text{ if $\theta \notin \Theta_{2\epsilon}$,} \\
0 & \text{ if $\theta \in \Theta_\epsilon$.}
\end{cases}
\end{equation}
Consider the set $J_{N,\epsilon}$ of $j$,
for which $\theta_j \notin \Theta_{2\epsilon}$.
The corresponding set sequence $\vecJ_\epsilon=\{J_{N,\epsilon}\}_{N\in\scrI}$
has, in view of Weyl's law (Theorem \ref{weyls}), density
$\Delta(\vecJ_\epsilon)=\mu(\scrD_0)+O(\epsilon)$, where the implied constant
is independent of $K$, since the measure of $\Theta_{2\epsilon}$ is at most
\begin{equation}
\sum_{\nu=1}^K \sum_{k=0}^{n_\nu-1} \frac{4\epsilon}{K n_\nu} = 4\epsilon.
\end{equation}
Now,
\begin{align}
\sum_{j\in J_{N,\epsilon}}
\left| \langle \Op_N(a) \varphi_j , \varphi_j \rangle \right|^2
& \leq
\sum_{j=1}^N h(\theta_j)
\left| \langle \Op_N(a) \varphi_j , \varphi_j \rangle \right|^2 \\
& \leq
\sum_{j=1}^N h(\theta_j)
\left\| \Op_N(a) \varphi_j \right\|^2 \\
& =
\sum_{j=1}^N h(\theta_j)
\langle \Op_N(|a|^2) \varphi_j , \varphi_j \rangle + o_{\epsilon}(N) .
\end{align}
Hence, by Theorem \ref{weyls2},
\begin{equation}\label{sososo}
\limsup_{N\to\infty} \frac1N \sum_{j\in J_{N,\epsilon}}
\left| \langle \Op_N(a) \varphi_j , \varphi_j \rangle \right|^2
\leq
\sum_{\nu=0}^\infty \int_{\scrD_\nu} |a|^2 d\mu \;
\int_0^1 h(\theta)\, \rho_\nu(\theta)\, d\theta .
\end{equation}
Now, under assumption (i) of the theorem,
\begin{equation}
\int_{\scrD_0} |a|^2 d\mu = 0,
\end{equation}
and furthermore
\begin{equation}
\sum_{\nu=1}^K \int_{\scrD_\nu} |a|^2 d\mu \;
\int_0^1 h(\theta)\, \rho_\nu(\theta)\, d\theta
=0
\end{equation}
since $h$ is supported outside the support of $\rho_1,\ldots,\rho_K$.
For the remaining sum,
\begin{equation}
\left| \sum_{\nu=K+1}^\infty \int_{\scrD_\nu} |a|^2 d\mu \;
\int_0^1 h(\theta)\, \rho_\nu(\theta)\, d\theta \right|
\leq \left| \sum_{\nu=K+1}^\infty \int_{\scrD_\nu} |a|^2 d\mu \right|
\leq \epsilon \max|a|^2
\end{equation}
by \eqref{sixfour}. Hence
\begin{equation}
\limsup_{N\to\infty} \frac1N \sum_{j\in J_{N,\epsilon}}
\left| \langle \Op_N(a) \varphi_j , \varphi_j \rangle \right|^2
= O(\epsilon)
\end{equation}
with $\epsilon>0$ arbitrarily small. Therefore there is a sequence
of values $\epsilon=\epsilon_N$ such that $\epsilon_N\to 0$ as $N\to\infty$,
and
\begin{equation}
\limsup_{N\to\infty} \frac1N \sum_{j\in J_N}
\left| \langle \Op_N(a) \varphi_j , \varphi_j \rangle \right|^2
= 0
\end{equation}
where $J_N:=J_{N,\epsilon_N}$. The proof for case (i)
is complete.

As to case (ii),
define $J_{N,\epsilon}'$ as the set of $j$,
for which $\theta_j \in \Theta_\epsilon$.
The corresponding set sequence
$\vecJ_\epsilon'=\{J_{N,\epsilon}'\}_{N\in\scrI}$
has density
$\Delta(\vecJ'_\epsilon)=1-\mu(\scrD_0)+O(\epsilon)$,
recall Weyl's law (Theorem \ref{weyls}).
Then, analogous to \eqref{sososo},
\begin{equation}
\limsup_{N\to\infty} \frac1N \sum_{j\in J_{N,\epsilon}'}
\left| \langle \Op_N(a) \varphi_j , \varphi_j \rangle \right|^2
\leq
\sum_{\nu=0}^\infty \int_{\scrD_\nu} |a|^2 d\mu \;
\int_0^1 [1-h(\theta)]\, \rho_\nu(\theta)\, d\theta  ,
\end{equation}
where all but the $\nu=0$ term vanish, since $a|_{\scrM-\scrD_0}=0$.
So
\begin{equation}
\limsup_{N\to\infty} \frac1N \sum_{j\in J_{N,\epsilon}'}
\left| \langle \Op_N(a) \varphi_j , \varphi_j \rangle \right|^2
\leq \int_{\scrD_0} |a|^2 d\mu \;
\int_0^1 [1-h(\theta)]\, d\theta
= O(\epsilon).
\end{equation}
\end{proof}

\begin{cor} \label{loccor}
There are set sequences $\vecI:=\{I_N\}_{N\in\scrI}$
and $\vecI':=\{I_N'\}_{N\in\scrI}$ with densities
\begin{equation}
\Delta(\vecI)=\mu(\scrD_0), \qquad
\Delta(\vecI')=1-\mu(\scrD_0),
\end{equation}
such that
\begin{itemize}
\item[(i)]
the eigenstates $\varphi_j$, $j\in I_N$, are semiclassically
localised in $\scrD_0$;
\item[(ii)]
the eigenstates $\varphi_j$, $j\in I'_N$, are semiclassically
localised in $\scrM-\scrD_0$.
\end{itemize}
\end{cor}

\begin{proof}
This is a straightforward consequence of Theorem \ref{locthm}
and Chebyshev's inequality.
\end{proof}

\begin{thm} \label{locthm2}
If in addition to the assumptions of Theorem \ref{locthm}
the phases $\alpha_\nu$ ($\nu=1,2,3,\ldots$) are linearly independent
over $\QQ$, then there are set sequences $\vecJ^\nu:=\{J_N^\nu\}_{N\in\scrI}$
with densities
\begin{equation}
\Delta(\vecJ^\nu)=\mu(\scrD_\nu)
\end{equation}
such that for any $a\in\C^\infty(\scrM)$  with $a|_{\scrD_\nu}=0$,
\begin{equation}
\lim_{N\to\infty} \frac1N \sum_{j\in J^\nu_N}
\left| \langle \Op_N(a) \varphi_j , \varphi_j \rangle \right|^2 = 0 .
\end{equation}
\end{thm}

\begin{proof}
This follows from a slight modification of the proof of Theorem \ref{locthm},
where the set $J^\nu_N$ is approximated by the set $J^\nu_{N,\epsilon}$
comprising those $j$ for which
\begin{equation}
\theta_j \in  \bigcup_{k=0}^{n_\nu-1}
\left[\frac{k}{n_\nu}+\alpha_{\scrD_\nu}-\frac{\epsilon \delta}{K n_\nu},
\frac{k}{n_\nu}+\alpha_{\scrD_\nu}+\frac{\epsilon\delta}{Kn_\nu} \right]
\mod 1 .
\end{equation}
The crucial observation is that, by the linear independence of the $\alpha_\nu$
over $\QQ$, there exists a $\delta=\delta_{\epsilon,K}>0$ small enough,
such that sets corresponding to different $\nu=1,2,3,\ldots,K$ are disjoint.
\end{proof}

Thus, by Chebyshev's inequality,
for a subsequence of $j\in J_N^\nu$ of density $\mu(\scrD_\nu)$
the eigenstates $\varphi_j$
are semiclassically localised in $\scrD_\nu$.

\section{Quantum ergodicity}\label{secQua}

Let us now turn to the question of quantum ergodicity for maps
which have one ergodic component $\scrD_0$ and are periodic on a remaining
countable collection $\scrD_1,\scrD_2,\ldots$ of domains.
Examples of linked twist maps with this property are discussed in Section
\ref{appErgodic}.

\begin{thm}\label{qethm}
Choose $\Phi$ and $U_N(\Phi)$ as in Theorem \ref{weyls}, and
suppose $\Phi$ acts ergodically on $\scrD_0$.
Let
$\varphi_1,\ldots,\varphi_N\in\CC^N$ be
an orthonormal basis of eigenstates of $U_N(\phi)$.
Then, for any $a\in\C^\infty(\scrM)$,
\begin{equation}\label{shsum}
\lim_{N\to\infty} \frac{1}{N} \sum_{j\in J_N}
\left| \langle \Op_N(a) \varphi_j , \varphi_j \rangle
-\int_{\scrD_0} a\, d\mu \right|^2 = 0.
\end{equation}
\end{thm}

\begin{proof}
We may assume without loss of generality that $\int_{\scrD_0} a\, d\mu = 0$
and $|a|\leq 1$. It is then sufficient to show\footnote{The
argument presented here is inspired by the proof of quantum ergodicity
for cat maps \cite{DeBievre01,Rudnick01}, cf.~also \cite{Zelditch96}.}
\begin{equation}
S_2(a,N):=\frac{1}{N} \sum_{j\in J_N}
\left| \langle \Op_N(a) \varphi_j , \varphi_j \rangle \right|^2 \to 0
\end{equation}
as $N\to\infty$.

For any given $T\geq 1$, we may write
\begin{equation}
a=a_T + a_T' + a_T^{\text{bad}}
\end{equation}
where
\begin{itemize}
\item[(i)]
$a_T\in\C^{\infty}$ has compact support contained in
$\scrD_0$ and the domain of continuity of $\Phi,\Phi^2,\ldots,\Phi^T$,
and furthermore $\int a_T\, d\mu = 0$, $|a_T|\leq 1$;
\item[(ii)]
$a_T'\in\C^{\infty}$ is supported inside $\scrM-\scrD_0$, and $|a_T'|\leq 1$;
\item[(iii)]
$\int_{\scrM} |a_T^{\text{bad}}|^2 \, d\mu  < T^{-1}$.
\end{itemize}
By the triangle inequality,
\begin{equation}
S_2(a,N)^{1/2} \leq S_2(a_T,N)^{1/2} + S_2(a_T',N)^{1/2}
+ S_2(a_T^{\text{bad}},N)^{1/2}
\end{equation}
By Theorem \ref{locthm},
\begin{equation}
\lim_{N\to\infty} S_2(a_T',N)= 0 .
\end{equation}
Furthermore, by the Cauchy-Schwartz inequality,
\begin{align}
S_2(a_T^{\text{bad}},N)
& \leq
\frac{1}{N} \sum_{j=1}^N
\left\| \Op_N(a_T^{\text{bad}}) \varphi_j  \right\|^2  \\
&  =
\frac{1}{N} \Tr \Op_N(|a_T^{\text{bad}}|^2)  + o_T(1)
\end{align}
and hence
\begin{equation}
\limsup_{N\to\infty} S_2(a_T^{\text{bad}},N) < T^{-1} .
\end{equation}
As to the remaining term,
\begin{equation} \label{thisterm}
S_2(a_T,N) =
\frac{1}{N} \sum_{j\in J_N}
|\langle \Op_N(a_T) \varphi_j,\varphi_j\rangle|^2 ,
\end{equation}
it remains to be proved that the limsup of \eqref{thisterm}
can be made arbitrarily small for sufficiently large $T$.
To this end define the ergodic average of $a_T$ by
\begin{equation}
a_T^T := \frac1T \sum_{n=1}^T a_T \circ \Phi^n .
\end{equation}
Since $\varphi_j$ are the eigenfunctions of $U_N(\Phi)$ we have
\begin{align}
S_2(a_T,N)
& =
\frac{1}{N} \sum_{j\in J_N}
\bigg|\frac1T \sum_{n=1}^T
\big\langle  U_N(\Phi)^{-n} \Op_N(a_T) U_N(\Phi)^{n}
\varphi_j,\varphi_j \big\rangle  \bigg|^2 \\
& \leq
\frac{1}{N} \sum_{j=1}^N
\bigg|\frac1T \sum_{n=1}^T
\big\langle  U_N(\Phi)^{-n} \Op_N(a_T) U_N(\Phi)^{n}
\varphi_j,\varphi_j \big\rangle  \bigg|^2 \\
& \leq
\frac{1}{N} \sum_{j=1}^N
\bigg\| \frac1T \sum_{n=1}^T U_N(\Phi)^{-n} \Op_N(a_T) U_N(\Phi)^{n}
\varphi_j  \bigg\|^2 \\
& = \frac{1}{N} \sum_{j=1}^N
\left\| \Op_N(a_T^T) \varphi_j \right\|^2 + o_T(1) ,
\end{align}
by Axiom \ref{corresp}.
Now
\begin{align}
\frac{1}{N} \sum_{j=1}^N \left\| \Op_N(a_T^T) \varphi_j \right\|^2
& = \frac{1}{N} \sum_{j=1}^N
\langle \Op_N(a_T^T)^\dagger \Op_N(a_T^T) \varphi_j , \varphi_j \rangle \\
& = \frac{1}{N} \sum_{j=1}^N
\langle \Op_N(|a_T^T|^2) \varphi_j , \varphi_j \rangle + o_T(1)\\
&= \int_{\scrD_0} |a_T^T|^2 d\mu + o_T(1).
\end{align}
We have
\begin{equation}
\left(\int_{\scrD_0} |a_T^T|^2 d\mu \right)^{1/2}
\leq \left(\int_{\scrD_0} |a^T|^2 d\mu \right)^{1/2}
+ \left(\int_{\scrD_0} |a_T^{\text{bad}T}|^2 d\mu \right)^{1/2}
\end{equation}
where
\begin{equation}
a^T := \frac1T \sum_{n=1}^T a \circ \Phi^n , \qquad
a_T^{\text{bad}T} := \frac1T \sum_{n=1}^T a_T^{\text{bad}} \circ \Phi^n .
\end{equation}
We have by Cauchy-Schwartz
\begin{align}
\int_{\scrD_0} |a_T^{\text{bad}T}|^2 d\mu
& =
\frac{1}{T^2} \sum_{j,k=1}^T  \int_{\scrD_0}  (a_T^{\text{bad}}\circ\Phi^j)
\overline{(a_T^{\text{bad}}\circ\Phi^k)}\, d\mu \\
& \leq
\frac{1}{T^2} \sum_{j,k=1}^T
\left(\int_{\scrD_0}  |a_T^{\text{bad}}\circ\Phi^j|^2\, d\mu \right)^{1/2}
\left(\int_{\scrD_0}  |a_T^{\text{bad}}\circ\Phi^k|^2\, d\mu \right)^{1/2}\\
& =
\int_{\scrD_0}  |a_T^{\text{bad}}|^2\, d\mu < \frac{1}{T},
\end{align}
using the $\Phi$-invariance of $\mu$ and assumption (iii).
Therefore
\begin{equation}
\limsup_{N\to\infty} S_2(a_T,N) \leq \int_{\scrD_0} |a^T|^2 d\mu +O(T^{-1}).
\end{equation}
Since $\Phi$ acts ergodically on $\scrD_0$,
we have a mean ergodic theorem for test functions $a\in\L^2(\scrM)$, i.e.,
\begin{equation}
\lim_{T\to\infty} \int_{\scrD_0} |a^T|^2 d\mu = 0 ,
\end{equation}
and hence $\limsup_{N\to\infty} S_2(a_T,N)$ becomes arbitrarily small
for $T$ sufficiently large.
\end{proof}

\begin{cor} \label{cor}
There is a set sequence $\vecI:=\{I_N\}_{N\in\scrI}$ with density
$\Delta(\vecI)=\mu(\scrD_0)$ such that
\begin{equation}
\langle \Op_N(a) \varphi_j , \varphi_j \rangle
\to \int_{\scrD_0} a\, d\mu
\end{equation}
for all $j\in I_N$, $N\to\infty$.
\end{cor}

\begin{proof}
Apply Chebyshev's inequality with the variance given in
\eqref{shsum}.
\end{proof}

\begin{appendix}
\section{\label{CVE} Converse quantum ergodicity. By Steve Zelditch}

The purpose of this appendix is to briefly make  some connections
between   the foregoing article of Marklof and O'Keefe and some
results in our articles \cite{Zelditch92, Zelditch94}
 concerning the distribution of eigenfunctions and eigenvalues on certain  Riemannian
 manifolds
 with an  open invariant component where
 the geodesic flow is periodic.

At the same time, we wish to emphasize the relevance of all of the
results to the  {\it converse quantum ergodicity problem}: are
quantum ergodic systems necessarily classically ergodic? As
discussed in the preceding article, it is  well-known that
classical ergodicity implies quantum ergodicity. However, there
are few results in the converse direction. There could exist
non-ergodic classical systems with quantum ergodic quantizations,
because the invariant sets might not be quantizable in any
suitable sense.

When the system is periodic in an open invariant set, one expects
to be able to quantize this feature and  prove that the system
cannot be quantum ergodic.  Examples where this has been carried
out  are given in the preceding article and in  \cite{Zelditch94}
(see also \cite{Schubert01}). Aside from fully integrable systems,
these appear to be the only examples where quantizations of
classically non-ergodic systems have been proved to be non-quantum
ergodic. It is very plausible that many other (if not all)
classically non-ergodic systems are quantum non-ergodic, e.g.
quantizations of KAM systems, but no rigorous proofs of this exist
at this time except in the partially periodic case.

The examples studied in \cite{Zelditch94} were special Riemannian
manifolds which we might informally call `pimpled spheres': We
take the  standard sphere $(S^n, g_0)$ and deform the metric, and
possibly the topology,  in a polar cap $B_r(x_0)$ of some small
radius $r$ to obtain a new Riemannian manifold $(M, g)$. We assume
that $M_r : = M \backslash B_r(x_0) \equiv S^n_r : = S^n
\backslash B_r(x_0)$ as Riemannian manifolds.  We then denote by
$\mcal_r \subset S^*M$ the smooth  manifold with boundary formed
by the closed geodesics of $(S^n, g_0)$ which lie in $M_r$. The
boundary consists of closed geodesics which intersect $\partial
B_r(x_0)$ tangentially. Thus, the boundary between the invariant
periodic component and the remaining   component is smooth and in
fact the geodesic flow is `almost-clean' in the sense of
\cite{Zelditch92}. This condition resembles (but is more
restrictive than) the Minkowski content zero condition on the
boundaries of the mixed phase space components in the preceding
paper.  Also, we did not fix the metric or the type of dynamics of
the geodesic flow  in the non-periodic component. In
\cite{Zelditch92} Theorem 3.20, we used the method of moments to
determine the Szeg\"o limit measure of the wave group in these
examples, i.e. the limit measure given in Marklof-O'Keefe's Theorem \ref{weyls}. 
As in the their paper, we used these results
to obtain eigenfunction distribution results. The following is a
corollary  of  Theorem A (b) of proved in \cite{Zelditch94}:

\begin{prop} The Laplacian $\Delta$ of a pimpled sphere $(M, g)$ is never quantum
ergodic. \end{prop}

Let us sketch the proof combining the argument of
\cite{Zelditch94} and that   of the foregoing paper. We assume the
reader is familiar with pseudodifferential and Fourier integral
operators, which are the observables and quantum maps in the
Riemannian setting.

\begin{proof} Let $A $ be a zeroth order pseudodifferential
operator on $M$ with essential support in $M_r$, i.e. assume the
complete symbol of $A$ is supported in $\mcal_r$. Since the
geodesic flow of $(M, g)$  is periodic of period $2 \pi$ in
$\mcal_r$,  $A \e^{2 \pi \i \sqrt{\Delta}}$ is a zeroth order
pseudodifferential operator on $M$. Using the calculation of the
symbol of the wave group in \cite{Duistermaat75}, it is simple to
see that the principal symbol of $A \e^{2 \pi \i \sqrt{\Delta}}$
equals $\e^{\i \alpha} \sigma_A$ for a constant $\e^{\i \alpha}$ (a
Maslov phase). Here, $\sigma_A$ is the principal symbol of $A$.
Thus, $\e^{2 \pi \i \sqrt{\Delta}}$ satisfies the hypothesis
\eqref{spezi} of the preceding article.

Now argue by contradiction. If $\Delta$ were quantum ergodic, we
would have 
\begin{equation}
\langle B \phi_j, \phi_j \rangle \to \int_{S^*M}
\sigma_B d\lcal
\end{equation} 
along a subsequence of density one of any
orthonormal basis  $\{\phi_j\}$ of $\Delta$-eigenfunctions. Here,
$\lcal$ is normalized Liouville measure and $B$ is a zeroth order
pseudodifferential operator.  Letting $B = A \e^{2 \pi \i
\sqrt{\Delta}}$,  we would obtain
\begin{equation}
\langle A \e^{2 \pi \i \sqrt{\Delta}} \phi_j, \phi_j \rangle \to \e^{\i \alpha}  \int_{S^*M}
\sigma_A d\lcal.
\end{equation}  
However, we also have
\begin{equation}
\langle A \e^{2 \pi \i \sqrt{\Delta}} \phi_j, \phi_j \rangle =
\e^{2 \pi \i \lambda_j} \langle A  \phi_j, \phi_j \rangle \sim \e^{2
\pi \i \lambda_j} \int_{S^*M} \sigma_A d\lcal. 
\end{equation} 
Here,
$\{\lambda_j\}$ are the eigenvalues of $\sqrt{\Delta}$ associated
to $\phi_j$.  It follows that $\e^{2 \pi \i \lambda_j} \to \e^{\i
\alpha}$  along a subsequence of eigenvalues of density one.

But this implies that all of the eigenvalues of $\e^{2 \pi
\i \sqrt{\Delta}}$ cluster around $\e^{\i \alpha}$, i.e. that
\begin{equation}
d\mu_{\lambda} :=  \frac{1}{N(\lambda)} \sum_{j: \lambda_j <
\lambda} \delta_{\e^{2 \pi \i \lambda_j}} \to \delta_{\e^{\i \alpha}}.
\end{equation} 
Here, $N(\lambda) = \# \{j : \lambda_j < \lambda\}.$ But the
weak limit of $d \mu_{\lambda}$ was calculated in
\cite{Zelditch92}, Theorem 3.20 (cf.~also 
Theorem \ref{weyls} in the preceding article),  and shown to be 
\begin{equation}
\frac{\lcal
(\mcal_r)}{\lcal(S^*M)} \delta_{\e^{\i \alpha}} + \bigg(1 - \frac{\lcal
(\mcal_r)}{\lcal(S^*M)}\bigg) d \theta.
\end{equation} 
This contradiction proves that $(M, g)$ is not quantum ergodic.

\end{proof}

In comparison, the argument in  \cite{Zelditch94} used a trace
formula for $Tr A \Pi_{\lambda}$ where $\Pi_{\lambda}$ is the
orthogonal projection onto the span of $\phi_j$ with $\lambda_j
\leq \lambda$. The above argument based on individual elements
seems
 more vivid.

\end{appendix}

\end{document}